\documentclass[twocolumn,preprintnumbers,showpacs,amsmath,amssymb]{revtex4}

\usepackage{psfrag}
\usepackage{color}
\usepackage{graphicx}
\usepackage{dcolumn}
\usepackage{bm}
\usepackage{epsfig}
\usepackage[notcite,notref]{}
\usepackage{mathrsfs}

\begin{document}

\title{Sphere on a plane: Two-dimensional scattering from a finite curved region}

\author{James R. Anglin$^1$ and Etienne Wamba$^{1,2,3}$}

\affiliation{$^1$Physics Department and State Research Center OPTIMAS, University of Kaiserslautern,
Erwin-Schr\"odinger-Stra{\ss}e, D-67663 Kaiserslautern, Germany\\
$^2$Faculty of Engineering and Technology, University of Buea, P.O. Box 63, Buea, Cameroon\\
$^3$STIAS, Wallenberg Research Centre at Stellenbosch University, Stellenbosch 7600, South Africa}

\date{\today}

\begin{abstract}
Non-relativistic particles that are effectively confined to two dimensions can in general move on curved surfaces, allowing dynamical phenomena beyond what can be described with scalar potentials or even vector gauge fields. Here we consider a simple case of piecewise uniform curvature: a particle moves on a plane with a spherical extrusion. Depending on the latitude at which the sphere joins the plane, the extrusion can range from an infinitesimal bump to a nearly full sphere that just touches the plane. Free classical motion on this surface of piecewise uniform curvature follows geodesics that are independent of velocity, while quantum mechanical scattering depends on energy. We compare classical, semi-classical, and fully quantum problems, which are all exactly solvable, and show how semi-classical analysis explains the complex quantum differential cross section in terms of interference between two classical trajectories: the sphere on a plane acts as a kind of double slit.
\end{abstract}


\maketitle

\section{Introduction}
Considerable study has been given to the quantum dynamics of a non-relativistic particle confined to an arbitrary curved two-dimensional surface embedded in flat three-dimensional space. While it has been necessary even recently just to point out that in a curved space the Laplacian in the Schr\"odinger equation must become the appropriate Laplace-Beltrami operator\cite{BernardVoon}, much of the literature has concerned the addition of potential terms involving the intrinsic\cite{deWitt} or extrinsic\cite{JensenKoppe,daCosta,Liu,WangZong} curvature of the surface of motion, whether arising from dynamical confinement in the full three-dimensional dynamics \cite{JensenKoppe,daCosta,WangZong} or from considerations of operator ordering in constrained quantization \cite{deWitt,Liu,WangZong}. The theory has moved beyond these basic points: quantum motion in curved space has been given a phase space representation \cite{GFH}; vector potentials have been incorporated \cite{Oliveira}; and relativistic extensions have been provided, in the form of theories for Dirac electrons on curved surfaces of topological insulators \cite{TakaneImura} or even in wormhole geometries realized with graphene \cite{GonzalezHerrero}. Scattering of two-dimensional particles from curved regions has been analysed perturbatively in Born approximation \cite{OMA18}, including lattices of small surface bumps; perturbative scattering from curved surfaces with delta-function defects has been studied in \cite{BM19}. An experiment studying electrons on the curved inner surfaces of multi-electron bubbles in liquid helium has even been reported \cite{Bubbles}. 

Explicit exact solutions for quantum particle motion on a curved surface have only been provided in a few simple cases, however, including spheres\cite{Liu}, ellipsoids\cite{BernardVoon,Mladenov}, and tori\cite{WangZong}. Here we provide explicit solutions for a significantly different kind of curved surface, namely one formed by joining a portion of a sphere to an infinite plane, to make a bump or bubble that bulges out of the plane, which is otherwise flat. This `sphere-on-a-plane' surface thus has piecewise uniform curvature, namely zero (in the planar portion) and negative (on the spherical portion). Three examples of the kind of surface we mean are shown in Fig.~1. We will consider the finite curved portion of this surface---the spherical extrusion---as a scatterer, and compute its differential cross section: classically, semi-classically, and quantum mechanically. Our paper can thus be considered complementary to Ref.~\cite{OMA18}, inasmuch as we consider a single class of simple geometries rather than a generic bumpy surface, but go beyond Born approximation. 
\begin{figure}
\centering
\includegraphics[width=0.45\textwidth, trim=0mm 0mm 0mm 0mm ,clip]{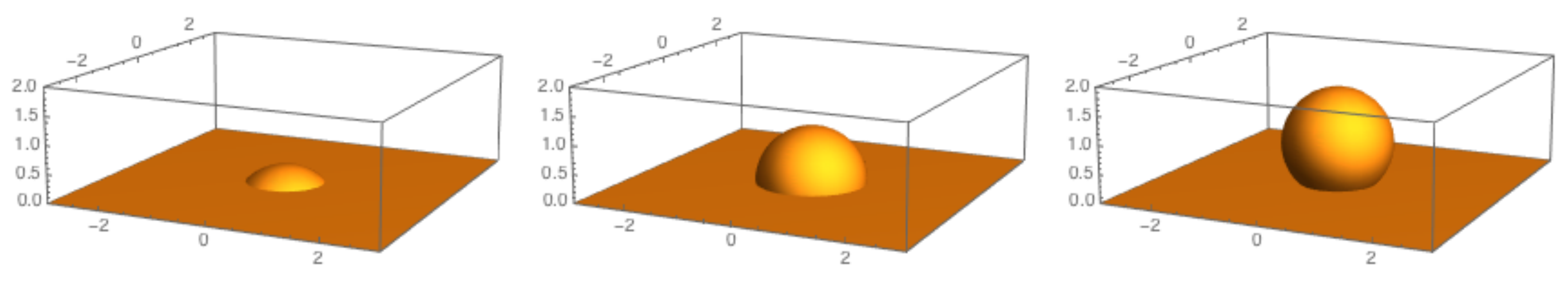}
\caption{Surfaces of piecewise uniform curvature: a portion of a sphere inserted into a plane. The radius of the sphere, and the latitude at which the sphere joins the plane, are independent parameters. We express this latitude as polar angle from the top of the sphere, denoted with $\alpha$. The examples shown here have $\alpha = \pi/4$, $\pi/2$, and $3\pi/4$, from left to right.}
\label{row}
\end{figure}

We do this because we wish to consider particle deflection by surface curvature as an analog to the specular reflection from hard walls which defines dynamical billiards\cite{Sinai,Heller}. Billiard models are good tools for examining the relationship between quantum and classical mechanics, because in billiard models the classical trajectories do not depend on energy. Any particle trajectory can be traversed at any energy, depending on how fast the particle travels the path. Even though the classical phase space is four-dimensional, therefore, one can describe the classical motion completely in terms of orbits in the two-dimensional position space, and then compare this to quantum wave functions in two-dimensional position space---and only the quantum energy needs to be considered as a varying parameter, because the classical paths are energy-independent. 

This convenient and instructive feature of billiard models is shared by curved surfaces, in which the classical motion is at arbitrary constant speed along geodesics that are the same for all speeds, and thus independent of particle energy. The ultimate interest in motion on curved surfaces in general may lie in complex geometries that provide chaotic motion as billiards can, but our contribution here will only be to consider the simplest case of scattering, in an infinite plane, from an inserted spherical extrusion. The billiard analog of our spherical extrusion would be a hard disk barrier. While the only thing a hard disk can do is to reflect particles specularly, the spherical extrusion can affect particle motion slightly, or greatly, depending on whether the spherical extrusion is only a tiny bump, or a nearly full sphere.

\subsection{Structure of the paper}
In Section II we will describe our geometry precisely and introduce our notation, and then solve the classical scattering problem, deriving a surprisingly compact exact expression for the classical differential cross section for spheres of arbitrary size and joining latitude. 

In Section III we will then briefly discuss the issue of whether the potentials applied to three-dimensional quantum particles, in order to constrain their low-energy motion to follow a two-dimensional surface, must always induce a non-constant effective potential within that surface. Although particular effective potentials have been derived in the literature, we will show that in general any potential whatever is possible, depending on exactly how the three-dimensional motion is constrained to a surface, and that it is therefore legitimate to consider the simplest case, in which the two-dimensional potential vanishes and two-dimensional dynamics involves only the intrinsic geometry of the surface. We will then exactly solve the quantum scattering problem for our sphere-on-a-plane geometry with zero potential.

In Section IV we will construct the semiclassical approximation for the sphere-on-a-plane scattering problem, and compare it to both our classical and quantum results. The semiclassical approximation to the differential cross section will turn out to be essentially the classical result, except with a kind of two-slit interference pattern superimposed on it. We will find the semiclassical approximation to be excellent whenever the quantum wavelength is shorter than the radius of the circle on which the sphere meets the plane. 

Finally in Section V we will briefly discuss our results and conclude.

\section{The Classical Problem}
\subsection{Motion on the sphere-in-a-plane}
\subsubsection{Notation and coordinates}
Our geometry will be described as follows. The radius of the spherical extrusion will be denoted $R$. The sphere is joined to the plane at polar angle $\alpha$, so that the radius in the plane of the joining circle is $R\sin\alpha$. See Fig.~2 for a side view of our extrusion geometry, which looks somewhat different for $\alpha<\pi/2$ and $\alpha>\pi/2$. The two cases of $\alpha<\pi/2$ and $\alpha>\pi/2$ will sometimes need to be considered separately in our derivations, but all our results will come out as unified formulas that apply in both regimes.
\begin{figure}
\centering
\includegraphics[width=0.45\textwidth, trim=0mm 0mm 0mm 0mm ,clip]{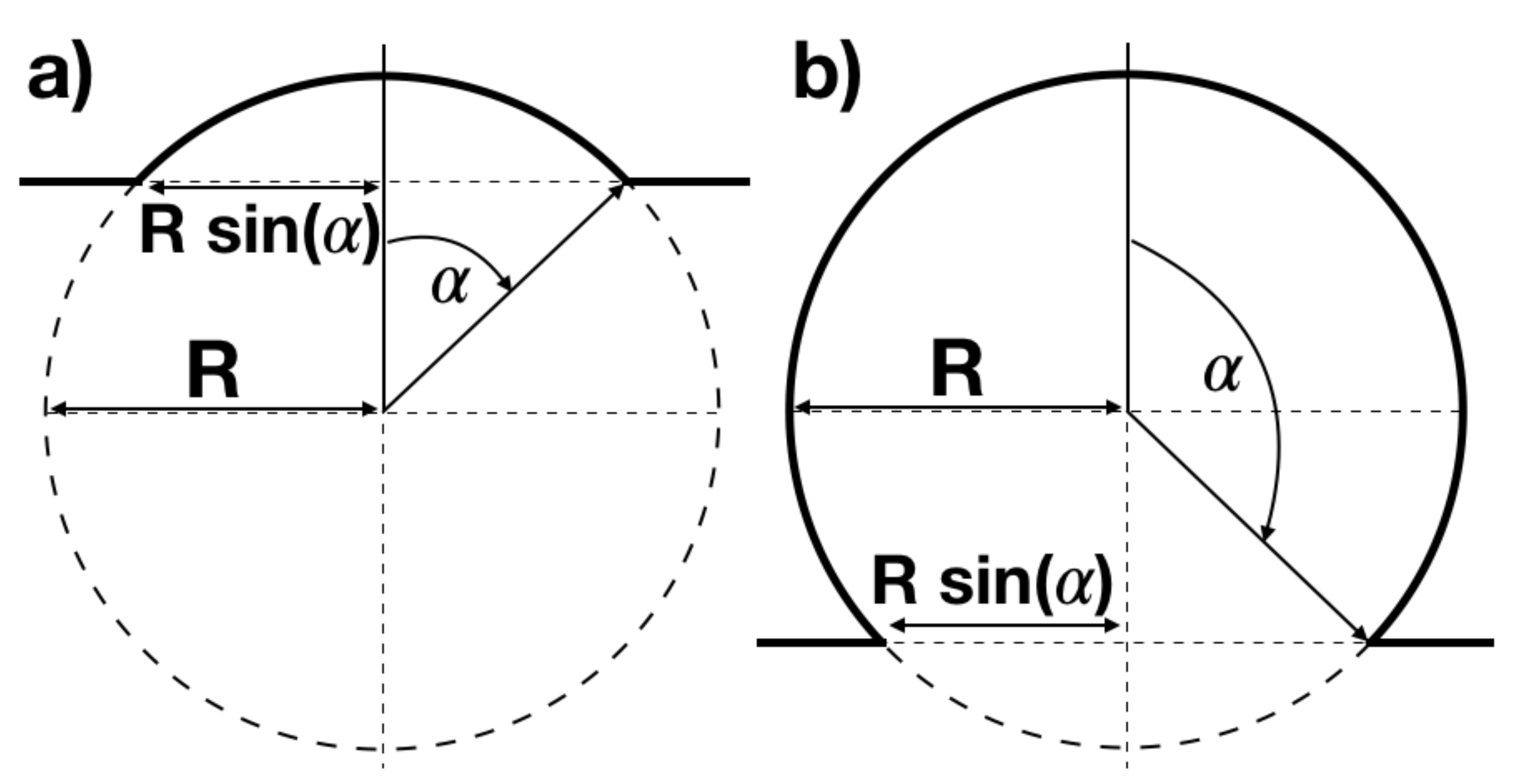}
\caption{View of our surface in section from the side. The radius of the sphere is $R$; it is joined to the plane at polar angle $\alpha$. For $\alpha<\pi/2$, as in the sub-figure a), the spherical part of the surface is a convex bump. For $\alpha>\pi/2$, as in sub-figure b), it is a more-than-hemispherical bubble. In both cases the radius of the junction circle, within the plane, is $R\sin\alpha$.} 
\label{side}
\end{figure}

The Lagrangian for a classical particle of mass $M$ on a two-dimensional surface is
\begin{equation}\label{Lagrange}
\mathscr{L} = \frac{M}{2}\sum_{i,j=1}^{2}g_{ij}\frac{d{r}_{i}}{dt}\frac{d{r}_{j}}{dt}
\end{equation}
where $r_{1,2}$ are arbitrary coordinates in the surface and $g_{ij}$ is the two-dimensional metric tensor for the surface in those coordinates. Since our problem has rotational symmetry around the center axis of the sphere, we will use dimensionless polar coordinates $\rho=r/R$ and $\phi$ such that the embedding of our surface in three-dimensional flat space is expressed in 3D Cartesian coordinates as
\begin{equation}\label{embed}
\left(\begin{array}{c}x\\ y\\ z\end{array}\right)=\left\{\begin{array}{lcl} \left(\begin{array}{c} R\cos\phi\,\sin\rho\\ R\sin\phi\,\sin\rho\\ R\cos\rho
\end{array}\right)&,& \rho<\alpha\\
\left(\begin{array}{c}R(\rho-\alpha+\sin\alpha)\cos\phi\\  
R(\rho-\alpha+\sin\alpha)\sin\phi\\ R\cos\alpha\end{array}\right)&,&\rho>\alpha\;.\end{array}\right.
\end{equation}
This implies the two-dimensional metric
\begin{equation}\label{metric}
\left(\begin{array}{cc}g_{\rho\rho}&g_{\rho\phi}\\ g_{\phi \rho}&g_{\phi\phi}\end{array}\right) = R^{2}\left(\begin{array}{cc}1&0\\ 0&g(\rho)\end{array}\right)
\end{equation}
where
\begin{equation}\label{grho}
g(\rho) = \left\{\begin{array}{lcl} \sin^{2}\rho&,&\rho<\alpha\\ (\rho-\alpha+\sin\alpha)^{2}&,&\rho>\alpha\end{array}\right.\;.
\end{equation}
We will use these coordinates and this metric throughout this paper; in this classical section we will also use the Cartesian $x,y$ coordinates for straight-line trajectories in the plane.

\subsubsection{Geodesics}
In this classical problem the trajectories on planes and spheres are well known: the geodesics are straight lines on the plane, and on the sphere they are great circles. If $\alpha\geq\pi/2$ there will exist some closed great circle orbits, such as the equator, which remain on the sphere and never enter the plane. There are likewise many trajectories in the plane that simply pass by the sphere without ever touching it. We are interested here, however, in trajectories that cross from the plane onto the sphere and then back into the plane. These must consist of half-lines and great circle arcs which connect on the contact circle; the non-trivial question is, Which half-lines and arcs must connect together to make up a total trajectory?

This problem can be solved with two-dimensional geometrical constructions, using the fact that great circles on the sphere become ellipses with semi-major axis $R$ when projected into the $x,y$ plane. It is both faster and more easily related to less symmetrical problems, however, to use the Euler-Lagrange equations of motion for the Lagrangian (\ref{Lagrange}). For the $\phi$ coordinate they yield
\begin{equation}\label{theta}
MR^{2}g(\rho)\frac{d\phi}{dt} = J
\end{equation}
for some constant $J$. By differentiating our embedding (\ref{embed}) with respect to $t$ we can confirm that $J$ is nothing but the angular momentum of the particle about the central axis of the sphere:
\begin{equation}\label{angmom}
J\equiv M[x(t)\frac{d}{dt}y(t)-y(t)\frac{d}{dt}x(t)]\;.
\end{equation}
This Cartesian form for $J$ is convenient for the straight-line trajectories in the plane.

The radial equation for $\rho$ has the energy as a first integral, allowing the usual reduction to a first-order equation:
\begin{equation}\label{energy}
\frac{d\rho}{dt} = \pm\frac{v}{R}\sqrt{1-\frac{J^{2}}{M^{2}v^{2}R^{2}g(\rho)}}
\end{equation}
for a constant $v>0$ that can be seen, by inserting (\ref{angmom}) and (\ref{energy}) in (\ref{embed}), to be the particle's constant speed. Every incident trajectory begins on the $-$ branch of the $\pm$, with $\rho$ decreasing monotonically towards its minimum value, where the trajectory makes its closest approach to the `north pole' of the sphere. There, where $\dot{\rho}$ is instantaneously zero, the branch changes from $-$ to $+$, and thereafter $\rho$ increases monotonically, as the particle exits the sphere and continues away in the plane to infinity. 

\subsubsection{Scattering trajectories}
In the plane, Eqns.~(\ref{theta}) and (\ref{energy}) are simply polar representations of straight lines, which can also be represented more simply in Cartesian terms. Without loss of generality we can take an initial straight-line trajectory which encounters the spherical extrusion from the negative $x$ direction, moving in the positive $x$ direction:
 \begin{equation}\label{incident}
 \left(\begin{array}{c}x(t)\\ y(t)\end{array}\right)_{\text{in}}=\left(\begin{array}{c}x(0)\\ b \end{array}\right)+vt\left(\begin{array}{c}1\\ 0\end{array}\right)\;,
\end{equation}
where $b$ is the \textit{impact parameter}. Unless $|b|<R\sin\alpha$ the incident particle will simply pass by the spherical extrusion without any scattering, so we can restrict our attention to these cases. For later convenience we will define the angle $\beta\in [-\pi/2,\pi/2]$ such that
\begin{equation}
b=R\sin\alpha\,\sin\beta\;.
\end{equation}
By construction, therefore, this incident particle trajectory meets the spherical extrusion at $\rho=\alpha$, $\phi = \pi-\beta$. 
The angular momentum $J$ of this trajectory is determined by the velocity $v$ and impact parameter $b$: it is easily found from (\ref{angmom}) that 
\begin{equation}\label{JMvb}
J=-Mvb.\end{equation}

Once the particle moves onto the spherical surface, it maintains its constant speed $v$ but follows the unique great circle on the sphere which (i) has angular momentum $J=-Mvb$ and (ii) meets the joining circle at $(\rho,\phi)=(\alpha,\pi-\beta)$. The exactly equivalent description of this motion in polar coordinates, less geometrically clear but more computationally convenient, is that the particle's radius $\rho$ will decrease from $\rho_{1}=\alpha$ (the $-$ branch in (\ref{energy})) until it reaches the turning point $\rho_{0}$ at which $d\rho/dt=0$, then change to the $+$ branch of (\ref{energy}) and increase until it again reaches $\rho=\alpha$, where the particle exits from the spherical extrusion and re-enters the plane. The particle's angle coordinate at this time, $\phi=\phi_{\mathrm{exit}}$, can be computed explicitly from 
\begin{eqnarray}
\phi_{\mathrm{exit}}&=&\pi-\beta + 2\int_{\alpha}^{\rho_{0}}\!d\rho\,\frac{d\phi/dt}{d\rho/dt}\nonumber\\
&=&\pi-\beta + 2\int_{\alpha}^{\rho_{0}}\!d\rho\,\frac{\sin\alpha\,\sin\beta}{\sin\rho\,\sqrt{\sin^{2}\rho-\sin^{2}\alpha\,\sin^{2}\beta}}\nonumber\\
&=&\pi-\beta + 2\tan^{-1}\left(\frac{\sin\alpha\,\sin\beta\,\cos\rho}{\sqrt{\sin^{2}\rho-\sin\alpha\,\sin\beta}}\right)\Big\vert_{\rho=\rho_{0}}^{\rho=\alpha}\nonumber\\
&=&\pi-\beta + 2\left(\tan^{-1}(\cos\alpha\,\tan\beta)-\frac{\pi}{2}\right)\nonumber\\
&=& 2\tan^{-1}(\cos\alpha\,\tan\beta)-\beta\;.
\end{eqnarray}
The same result can be reached from great circle geometry but the derivation is longer.

Returning now to Cartesian coordinates for the further motion in the plane, the point of exit from the sphere and reentry to the plane is
 \begin{equation}\label{entry}
 \left(\begin{array}{c}x\\ y\end{array}\right)_{\mathrm{reentry}}=R\sin\alpha\left(\begin{array}{c}\cos\phi_{\mathrm{exit}}\\ \sin\phi_{\mathrm{exit}} \end{array}\right)
\end{equation}
and therefore the final part of the trajectory which continues further into the plane and has the same energy constant $\Omega$, hence the same speed $v$, must be
 \begin{equation}\label{continue}
 \left(\begin{array}{c}x(t)\\ y(t)\end{array}\right)_{\text{out}}=R\sin\alpha\left(\begin{array}{c}\cos\phi_{\mathrm{exit}}\\ \sin\phi_{\mathrm{exit}} \end{array}\right)+v(t-t_{\mathrm{exit}})\left(\begin{array}{c}\cos\theta\\ \sin\theta \end{array}\right)
\end{equation}
for some scattering angle $\theta$. 

The angular momentum (\ref{angmom}) of this exiting trajectory must also be the same $J=-Mvb=-MvR\sin\alpha\,\sin\beta$ of the incident trajectory. We therefore have
\begin{equation}\label{angmom2}
-\sin\beta=\cos\phi_{\mathrm{exit}}\,\sin\theta-\sin\phi_{\mathrm{exit}}\,\cos\theta \equiv \sin(\theta-\phi_{\mathrm{exit}})\;.
\end{equation}
One solution to this equation would be $\theta\to2\tan^{-1}(\cos\alpha\,\tan\beta)\pm \pi$, but this would imply $\dot\rho = -(v/R)\cos(\beta)<0$ at the moment of exit, when at this point we must be on the $+$ branch of (\ref{energy}). Hence the only scattering angle which satisfies the radial equation of motion as well as (\ref{angmom2}) is
\begin{equation}\label{thetabeta}
\theta = 2[\tan^{-1}(\cos\alpha\,\tan\beta)-\beta]\qquad \mathrm{mod}(2\pi)\;.
\end{equation}
For our use in Section IV, below, we recall that $\sin\beta = b/(R\sin\alpha)=-J/(MvR\sin\alpha)$, which means that $\sin\beta\to -m/(kR\sin\alpha)$ when in the quantum problem $J\to\hbar m$ and $Mv\to \hbar k$. This means that the classical scattering angle is a function of $\alpha$ and $J/(MvR)\to m/(kR)=:\mu$,
\begin{align}\label{ThetaC}
\theta &= \Theta\Big(\alpha,\frac{J}{MvR}\Big)\\
\Theta(\alpha,\mu) &= 2\sin^{-1}\Big(\frac{\mu}{\sin\alpha}\Big)-2\tan^{-1}\Big( \frac{\mu\cos\alpha}{\sqrt{\sin^2\alpha-\mu^2}}\Big)\;.\nonumber
\end{align}
The classical scattering angle $\theta$ will appear in the semi-classical theory of Section IV, in this form as $\Theta(\alpha,\mu)$.

\subsubsection{Scattering angle for given impact parameter}
We recall that $|\beta|\leq\pi/2$ by definition, because for scattering the impact parameter $b=R\sin\alpha\,\sin\beta$ must lie between $\pm R\sin\alpha$. If we vary the impact parameter from $-R\sin\alpha$ to $+R\sin\alpha$, $\tan\beta$ increases monotonically from $-\infty$ to $+\infty$. If $\alpha>\pi/2$, then $\cos\alpha\,\tan\beta$ \textit{decreases} monotonically from $+\infty$ to $-\infty$, and hence both $\tan^{-1}(\cos\alpha\,\tan\beta)$ and $-\beta$ decrease monotonically from $\pi/2$ to $-\pi/2$. This means that $\theta$ decreases from $2\pi$ to $-2\pi$, modulo $2\pi$. In other words, when the spherical extrusion is larger than a hemisphere, the scattering angle covers the full circle, so that every possible scattering angle, including directly backwards, occurs for some impact parameter. In fact $\theta$ actually sweeps through a range of $4\pi$, meaning that every scattering angle is obtained twice, for two different values of the impact parameter $b$.

If $\alpha<\pi/2$, on the other hand, so that the spherical extrusion is less than a hemisphere, then $\tan^{-1}(\cos\alpha\,\tan\beta)$ increases monotonically with $b$ while $-\beta$ decreases. As a result, $\theta$ does not change monotonically with $b$, but has an extremum when
\begin{eqnarray}
\frac{d\theta}{d\beta} &=& 2\left( \frac{\cos\alpha}{1-\sin^{2}\alpha\,\sin^{2}\beta}-1\right)\to 0\nonumber\\
&\Longrightarrow& \beta\to\pm\beta_{c}=\pm\sin^{-1}\left(\frac{1}{2\cos\frac{\alpha}{2}}\right)\nonumber\\
&\Longrightarrow& \theta\to\pm\theta_{c}=\pm 2\sin^{-1}\left(\tan^{2}\frac{\alpha}{2}\right)\;.
\end{eqnarray}
So if the extrusion is smaller than a hemisphere, no particles are scattered by more than the angle $\pm\theta_{c}$, and these scattering angles are caustic angles at which scattered trajectories ``pile up'' because the scattering angle is reversing direction as a function of impact parameter. Those scattering angles $|\theta|<\theta_{c}$ which are realized are again each realized twice, for different impact parameters.

\subsection{Differential cross sections}
While $\theta(\beta)$ is the natural solution to the particle motion as an initial value problem, the differential cross section $|db/d\theta|$ defines what infinitesimal range of impact parameters $b$ contributes to scattering within an infinitesimal range of scattering angles $\theta$. In this sense it is a final value problem instead of an initial value problem, but it is normally used to give probabilistic answers to a probabilistically posed initial value question:
if an ensemble of incident particles encounters the scatterer with an evenly distributed range of impact parameters, then $|db/d\theta|$ as a function of $\theta$ gives the probability density for scattering at the angle $\theta$. 

\subsubsection{Hard disk example}
\begin{figure}
\includegraphics[width=0.4\textwidth, trim=0mm 0mm 0mm 0mm,clip]{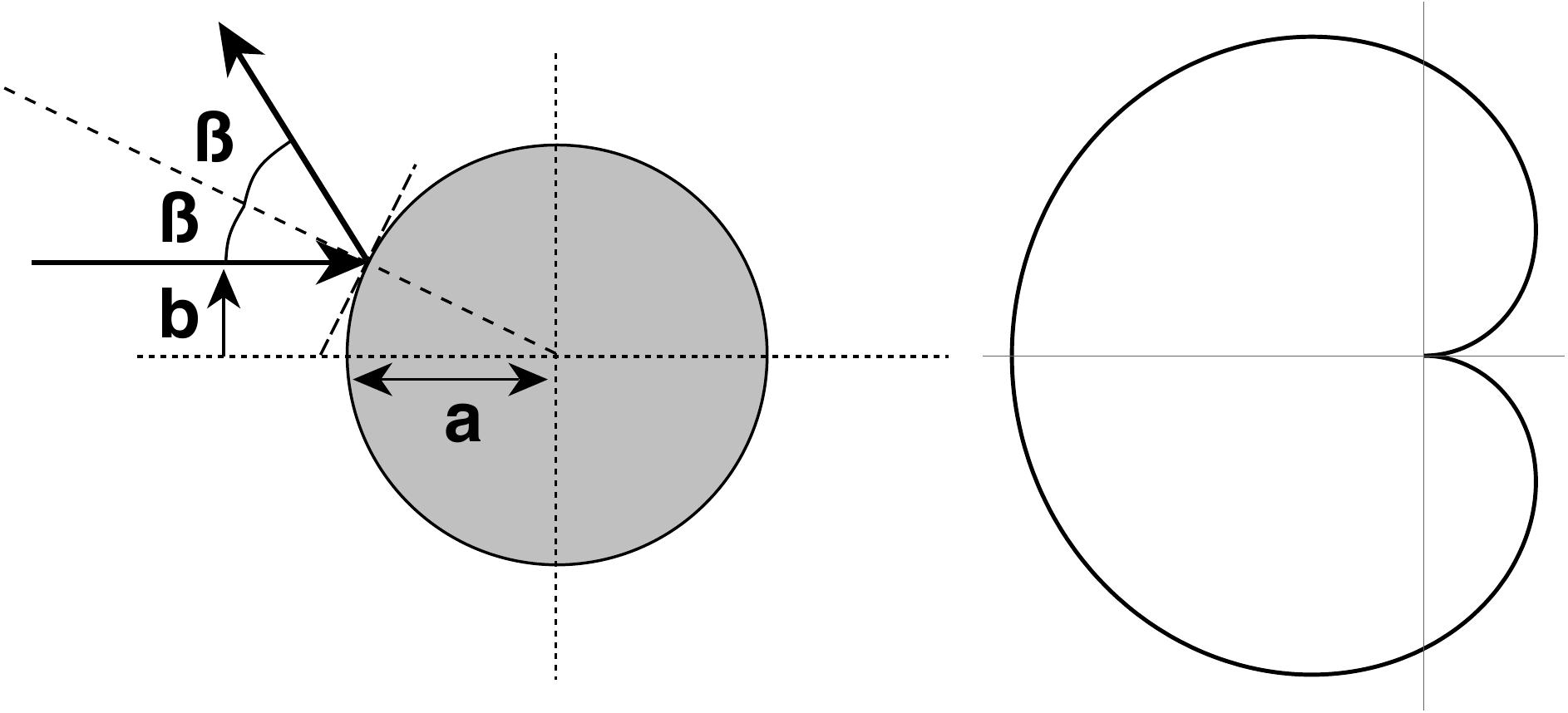}
\caption{Classical scattering from a hard disk of radius $a$. \textbf{Left:} a particle with impact parameter $b=a\sin\beta$ is reflected. \textbf{Right:} Polar plot of
the cardioid $r=(1/4)\sin(|\theta|/2)$, which is the differential cross section divided by the total cross section.}
\label{diskplot}
\end{figure}

Since differential cross sections may be more familiar in quantum mechanics than they are classically, we briefly review this classical scattering theory for the simple example of a hard disk scatterer of radius $a$. This geometry is shown in Fig.~\ref{diskplot}, and from the figure we can immediately see that for impact parameter $b=a\sin\beta > 0$ the exit angle is $\theta = \pi-2\beta$. For negative $b$, it is clear from the problem's vertical reflection symmetry that we must simply reverse the sign of $\theta$. It follows that in general
\begin{equation}
\theta = 
\left\{
	\begin{matrix}
\pi\,\mathrm{sgn}(\beta)-2\beta\;, & \beta \neq 0\\
\pi\; , & \beta = 0.
\end{matrix}
\right.
\end{equation}
To obtain the differential cross section, however, we must invert this relation and find $b=a\sin\beta$ as a function of $\theta$. For the hard disk this can be easily done: $\beta = (1/2)(\pi\,\mathrm{sgn}(\theta)-\theta)$, so that for $b=a\sin\beta$ we obtain
\begin{equation}
b(\theta) = a\,\mathrm{sgn}(\theta)\cos\frac{\theta}{2}\;.
\end{equation}
The differential cross section is then
\begin{equation}\label{dbdisk}
\Big\vert\frac{db}{d\theta}\Big\vert = \frac{a}{2}\sin\frac{|\theta|}{2}\;,
\end{equation}
which is the cardioid curve plotted in the right panel Fig.~\ref{diskplot}. The shape of the cardioid means that when an ensemble of particles with evenly distributed impact parameters strike the disk from the left, many are scattered nearly backward, while only few strike near enough to the edges of the disk to receive only slight deflections. The total cross section is identically equal to the disk diameter $2a$, because the integral of the differential cross section over the full range of scattering angles is by definition equal to the integral over all $b$ that produce any deflection, which is the range $-a<b<a$.

\subsubsection{Spherical extrusions}
To compute the classical differential cross section for our spherical extrusions, therefore, we need to invert our expression (\ref{thetabeta}) for $\theta(\beta)$ and obtain $b=R\sin\alpha\,\sin\beta$ as a function of $\theta$ instead.
Surprisingly, this can be done quite compactly, if we begin by using (\ref{thetabeta}) to compute
\begin{eqnarray}\label{sintrick}
\sin\frac{\theta}{2} &=& \frac{\cos\alpha\,\sin\beta-\sin\beta}{\sqrt{1+\cos^{2}\alpha\tan^{2}\beta}}\nonumber\\
&=& \sin\beta\,\cos\beta \frac{\cos\alpha - 1}{\sqrt{1-\sin^{2}\alpha\sin^{2}\beta}}\;.
\end{eqnarray}
Squaring both sides of (\ref{sintrick}) produces a quadratic equation for $\sin^{2}\beta$. Solving this and then applying trigonometric identities, and checking the results against (\ref{thetabeta}) itself to discard spurious roots produced by the squaring of  (\ref{sintrick}), eventually yields two solutions for $b=R\sin\alpha\,\sin\beta$ for each $\theta$, which we denote as $b(\theta)\to b_{\pm}(\theta)$:
\begin{eqnarray}\label{btheta}
b_{+}  &=& -R\,\mathrm{sgn}(\cos\alpha)\,\mathrm{sgn}(\theta) \cos^{2}\frac{\alpha}{2}\nonumber\\
&&\times \left(\sqrt{\left(1+\sin\frac{|\theta|}{2}\right)\left(\tan^{2}\frac{\alpha}{2}+\sin\frac{|\theta|}{2}\right)}\right.\nonumber\\
 &&\qquad+\left.\sqrt{\left(1-\sin\frac{|\theta|}{2}\right)\left(\tan^{2}\frac{\alpha}{2}-\sin\frac{|\theta|}{2}\right)}\right)\nonumber\\
b_{-}  &=& -R\,\mathrm{sgn}(\theta) \cos^{2}\frac{\alpha}{2}\\
&&\times\left(\sqrt{\left(1+\sin\frac{|\theta|}{2}\right)\left(\tan^{2}\frac{\alpha}{2}+\sin\frac{|\theta|}{2}\right)}\right.\nonumber\\
 &&\qquad-\left.\sqrt{\left(1-\sin\frac{|\theta|}{2}\right)\left(\tan^{2}\frac{\alpha}{2}-\sin\frac{|\theta|}{2}\right)}\right) \;.\nonumber
 \end{eqnarray}
 For $\alpha<\pi/2$, Eqn.~(\ref{btheta}) is only valid for $|\theta|<\theta_{c}=2\sin^{-1}(\tan^{2}\frac{\alpha}{2})$, because no scattering occurs with angles $|\theta|>\theta_{c}$. At the largest possible scattering angles, which are either $\pm\theta_{c}$ for $\alpha<\pi/2$ or $\pm\pi$ for $\alpha>\pi/2$, the two branches $b_{\pm}$ coincide. For $\theta\to0^{\pm}$, $b_{-}\to0$ while $b_{+}\to\mp R\,\mathrm{sgn}(\cos\alpha)\sin\alpha$. 
\begin{figure}
\includegraphics[width=0.45\textwidth, trim=0mm 0mm 0mm 0mm,clip]{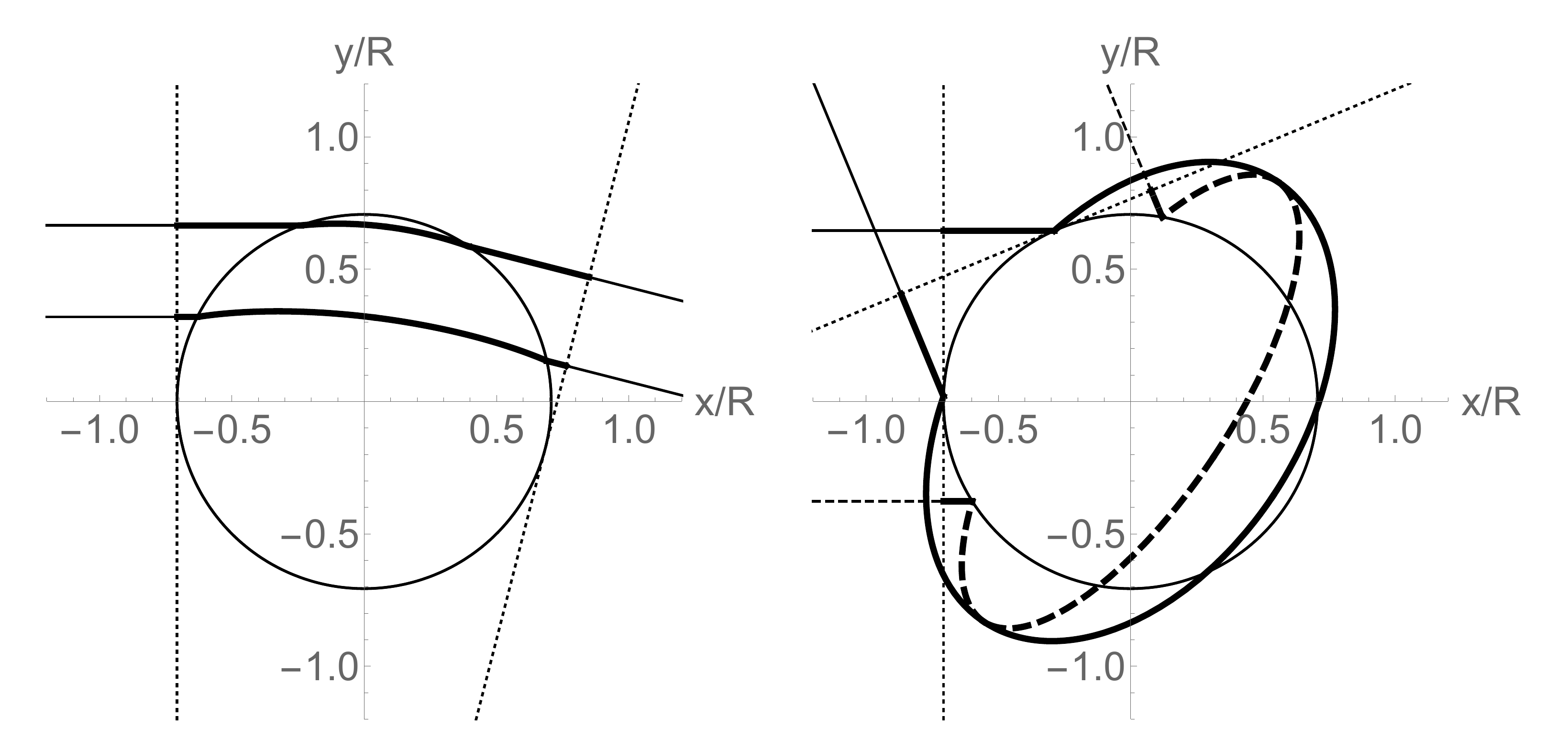}
\caption{Examples of pairs of trajectories with the same scattering angle $\theta$, for $\alpha=\pi/4$ and $\theta = 0.25$ (left panel) and $\alpha=3\pi/4$, $\theta=-5\pi/8$ (right panel). The trajectories are shown projected onto the plane, so that great circle arcs on the sphere appear as elliptical arcs. In each case the two trajectories enter the spherical extrusion from the left with two different impact parameters $b_\pm(\alpha,\theta)$, and exit at two different exit angles $\phi_\mathrm{exit}$, but emerge from the extrusion on parallel paths. One of the two trajectories in the right panel is shown dashed, to aid in distinguishing the two trajectories where their planar projections cross. Portions of the trajectories between the entry and exit tangents (dotted lines) are shown thicker, because the lengths of these thicker portions will appear in the semi-classical theory of Section IV.}
\label{classplot}
\end{figure}

Since both branches of $b_{\pm}$ contribute scattering in the $\theta$ direction, the differential cross section at $\theta$ has contributions from both: $|db/d\theta|=\sum_{\pm}|db_{\pm}/d\theta|$. It can be shown using trigonometric identities, or by plotting, that for $\alpha>\pi/2$, $b_{\pm}$ are both monotonically decreasing functions of $\theta$ over their full ranges of support in $\theta$, while for $\alpha<\pi/2$, $b_-$ is decreasing and $b_+$ is increasing. In the sum of the two branches, therefore, the first square root terms in each $b_{\pm}$ always cancel each other in the differential cross section, while the second square root terms add together. 

The final result for the classical differential cross section of the spherical extrusion is thus this surprisingly compact expression full of half-angles:
\begin{eqnarray}\label{classicalresult}
\Big\vert\frac{db}{d\theta}\Big\vert &=& 2R\sin\alpha\,D(\alpha,\theta)\\
D(\alpha,\theta)&=& \frac{\sqrt{1+\sin\frac{|\theta|}{2}}\left(1-2\cos^{2}\frac{\alpha}{2}\sin\frac{|\theta|}{2}\right)}{4\sin\alpha\sqrt{\tan^{2}\frac{\alpha}{2}-\sin\frac{|\theta|}{2}}}\;.\nonumber
\end{eqnarray}
The integral of this differential cross section over all scattering angles (\textit{i.e.} from $-\theta_{c}$ to $+\theta_{c}$ for $\alpha<\pi/2$ and from $-\pi$ to $+\pi$ for $\alpha>\pi/2$) is exactly $2R\sin\alpha$, as indeed it must be, because this is simply the full range of $b$ for which the particle touches the sphere and can be deflected. The dimensionless function $D(\alpha,\theta)$ thus gives the angular distribution of scattered particles, and is independent of $R$; it is plotted for a selection of different values of $\alpha$ in Fig.~\ref{classplot}. The area inside the closed curve defined as having radius $r(\theta)=D(\alpha,\theta)$ is identically 1 for all $\alpha$, if we count the caustic limits $\pm\theta_{c}$ as closing the curve for $\alpha<\pi/2$. The divergences of $D(\alpha,\theta)$ at the caustics are only inverse square roots $\sim (|\theta|-\theta_{c})^{-1/2}$, which are integrable.
\begin{figure}
\includegraphics[width=0.45\textwidth, trim=0mm 0mm 0mm 0mm,clip]{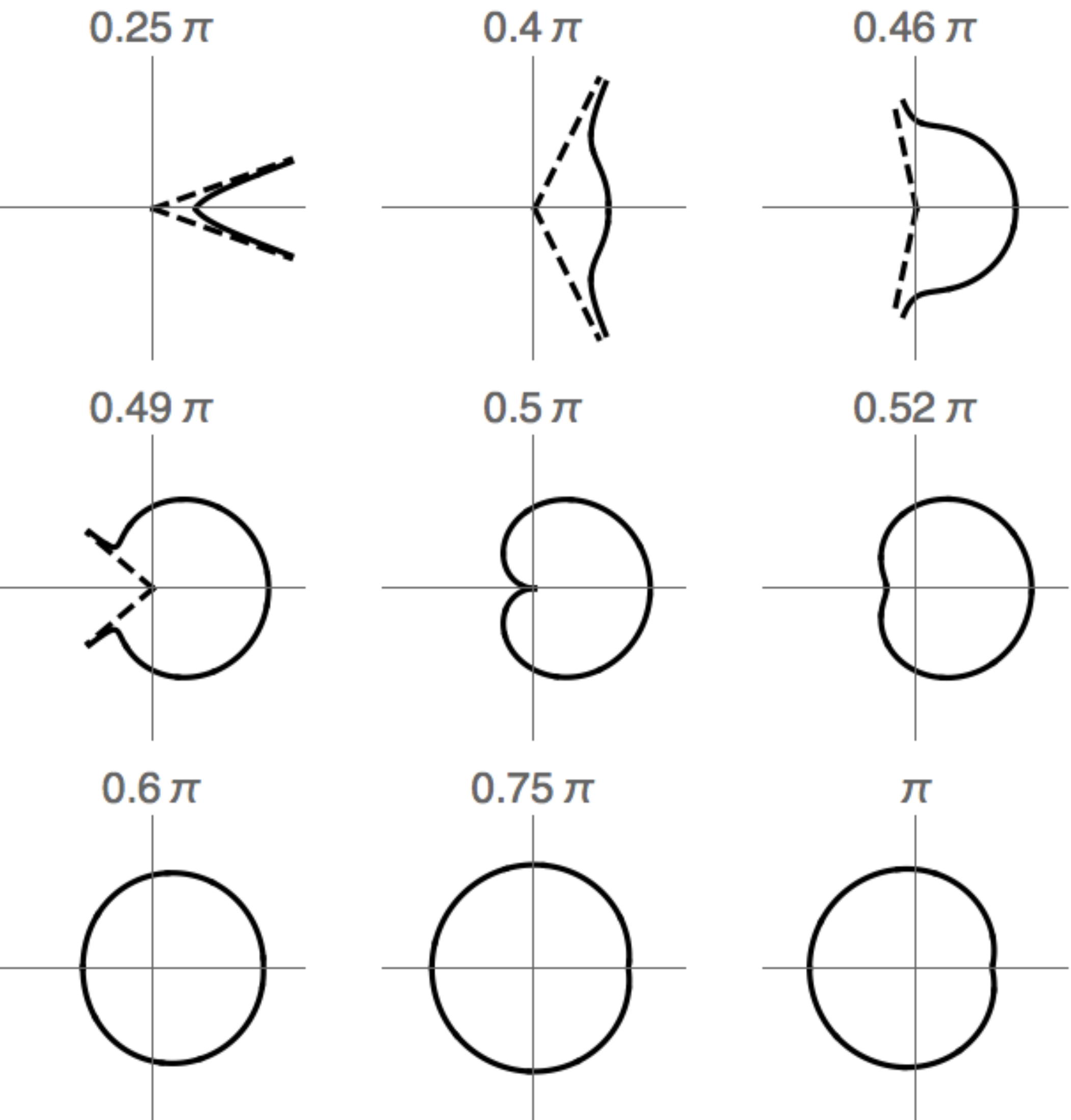}
\caption{Normalized differential cross section $D(\alpha,\theta)$ as polar plots in $\theta$ for the different values of $\alpha$ shown by the plot labels. To better show the shapes of the curves, the different plots are not all drawn to the same scale; the true size of the contours is determined by the fact that they all enclose unit area.}
\label{classplot}
\end{figure}

We can also note a few simple exact results for special cases of $\alpha$. For $\alpha\to0$ the maximum scattering angle $\theta_{c}\to0$ as well---a slight bump deflects only slightly---and so for all the very small angles into which scattering occurs we can write
\begin{equation}
\lim_{\alpha\to\epsilon^{+}}D(\alpha,\theta) = \frac{1/2}{\epsilon \sqrt{\epsilon^{2}-2|\theta|}}\;.
\end{equation}
Here $\epsilon $ is a small (positive) value of the parameter $\alpha$. If we let $\alpha\to0$ while increasing $R\sim 1/\alpha$ so that $a=R\sin\alpha$ remains fixed, then our spherical extrusion becomes a disk of finite radius $a$ with vanishing curvature. It nonetheless retains its finite total cross section $2a$---and this seems paradoxical, because in this limit of vanishing curvature the spherical extrusion should have no effect whatever on particle motion, since it is nothing but an arbitrarily designated circular region in the plane. The resolution of this paradox is in the definition of cross section as the measure of a set of trajectories that are affected by the spherical region. This definition takes no account of how greatly the trajectories are affected; any deflection at all is accounted as scattering, no matter how slight the deflection may be. When we take the limit of $R\to\infty$ and $\alpha\to0$ so that $R\sin\alpha$ stays constant, we approach the case where the sphere has no effect by keeping the same set of trajectories affected, but reducing the amount by which they are deflected.

For $\alpha\to\pi$ we instead have
\begin{equation}
D(\pi,\theta) = \frac{\sqrt{1+\sin\frac{|\theta|}{2}}}{8}\;.
\end{equation}
So if we let $R\to\infty$ while keeping $a=R\sin\alpha$ fixed by letting $\alpha$ approach $\pi$, then the finite contact circle of radius $a$ becomes the junction between the plane and a large spherical world into which particles can disappear from the plane for a long time, emerging eventually at some different angle from the one at which they entered, with a particular average distribution of exit angles given by the expression above, independent of $R$ (for fixed $a=R\sin\alpha$).

Finally, for the hemisphere we have 
\begin{equation}D(\frac{\pi}{2},\theta)=\frac{1}{4}\cos\frac{\theta}{2}\;,\end{equation}
which is exactly the same cardioid distribution of scattered particles that would be emitted by a hard disk, if the particles were incident from the opposite direction.

\section{The Quantum Problem}
\subsection{From three dimensions to two}
Any two-dimensional problem may be considered theoretically, but to restrict the three-dimensional motion of a physical particle to a two-dimensional surface, one must apply some confining force in the perpendicular direction. In Ref.~\cite{daCosta} it was concluded that the effective two-dimensional theory in quantum mechanics must therefore include a certain effective potential determined by the \textit{extrinsic} curvature of the surface. In fact this particular result for the effective potential is not a universal formula at all, however, but only the special case that applies if the strength of the perpendicular confinement is kept uniform over the two-dimensional surface. It is easy to see that even a small proportional variation of confinement strength can produce an arbitrary effective potential within the surface.

To confirm this it suffices to take the trivial case in which the two-dimensional surface to which three-dimensional motion should be confined is simply the plane $z=0$ in Cartesian coordinates. We let the confinement be enforced by a three-dimensional potential which is very slightly non-uniform over the plane:
\begin{equation}\label{3Dconfine}
V_{3D}(x,y,z) = \frac{\hbar^{2}z^2}{2M\lambda^{4}}\left[1+\epsilon^{2}\Phi\Bigl(\epsilon\frac{x}{\lambda},\epsilon\frac{y}{\lambda}\Bigr)\right]^{2}\;,
\end{equation}
where $M$ is the particle mass, $\lambda$ is the transverse confinement length, and $\epsilon\ll 1$. The effectively two-dimensional limit applies when we consider particle energies that are low in comparison to the transverse confinement scale,
\begin{eqnarray}
 E = [1+2\epsilon^{2}\mathcal{E}]\,\frac{\hbar^{2}}{2M\lambda^{2}}
 \end{eqnarray}
 for dimensionless $\mathcal{E} =\mathcal{O}(\epsilon^{0})$. 
 
 We can then introduce the scaled planar coordinates $(\tilde{x},\tilde{y})=(\epsilon/\lambda) (x,y)$ and the Ansatz\begin{widetext}
 \begin{equation}
 \Psi(x,yz) = e^{-\frac{z^{2}}{2\lambda^{2}}\left[1+\epsilon^{2}\Phi(\tilde{x},\tilde{y})\right]}\sum_{n=0}^{\infty}\epsilon^{4n}\psi_{n}\Bigl(\epsilon\frac{x}{\lambda},(\epsilon\frac{y}{\lambda}\Bigr)H_{2n}\Bigl(\frac{z\sqrt{1+\epsilon^{2}\Phi}}{\lambda}\Bigr)
\end{equation}
for the three-dimensional wave functions of energy eigenstates, with $H_{n}$ being the Hermite polynomials. The orthogonality of the Hermite polynomials then lets us read back out
\begin{eqnarray}\label{project}
\psi_{n}(\tilde{x},\tilde{y}) = \epsilon^{-4n}\int\!dz\, e^{-\frac{z^{2}}{2\lambda^{2}}[1+\epsilon^{2}\Phi(\tilde{x},\tilde{y})]} H_{2n}\Bigl(\frac{z\sqrt{1+\epsilon^{2}\Phi}}{\lambda}\Bigr) \Psi(x,y,z)\;.
\end{eqnarray}\end{widetext}
Inserting (\ref{project}) into the three-dimensional time-independent Schr\"odinger equation then reveals
\begin{eqnarray}\label{3DTISE}
\mathcal{E}\psi_{0} &=& -\frac{1}{2}\left[\frac{\partial^{2}}{\partial\tilde{x}^{2}}+\frac{\partial^{2}}{\partial\tilde{y}^{2}}\right]\psi_{0}+\frac{1}{2}\Phi(\tilde{x},\tilde{y})\psi_{0}+\mathcal{O}(\epsilon^{2})\;.\nonumber\\
\end{eqnarray}
Even a proportionally very small non-uniformity $\epsilon^{2}\Phi$ in strong perpendicular confinement will thus produce a significant effective potential in the two-dimensional low-energy dynamics.

For confinement to surfaces more complicated than a plane, Ref.~\cite{daCosta} has shown that the there is a non-trivial effective potential in two dimensions even when the perpendicular confinement strength is uniform. We now point out, however, that it may be a difficult matter in practice to achieve such uniform confinement that the result of \cite{daCosta} is valid---even when the surface to which motion should be confined is as simple as a plane or a cylinder. If the three-dimensional potential is complicated enough to confine the particle to a more complicated surface, then it will surely be more difficult still to make the confinement strength uniform. The particular potential derived in \cite{daCosta} must therefore be considered as a somewhat academic result. Quasi-two-dimensional motion that is realized in an experiment is likely to be subject to an effective potential that depends arbitrarily on the details of exactly how the confinement to two dimensions is achieved.

Conversely, however, it is in principle possible for practically any two-dimensional potential to be achieved, if the perpendicular confinement is implemented appropriately. The effects of local potentials on quantum motion in two dimensions have long been familiar, while the kinetic effects of spatial curvature, through a non-trivial Laplace-Beltrami operator, are less well understood. It is therefore a worthwhile theoretical contribution to analyze an informative model that involves only the Laplace-Beltrami operator, since this simple scenario is in principle no less realistic than the uniform confinement scenario assumed in Ref.~\cite{daCosta}. In this paper we will provide such a contribution, by assuming that the effective confinement from three dimensions to two has somehow been achieved in such a way as to leave no potential term in the effectively two-dimensional Schr\"odinger equation for quantum motion in the low-energy limit.

One result is easy to obtain immediately: introducing a curved region within an asymptotically flat plane that extends to infinity can never create any bound states. If there were such a bound state, then in the flat plane at infinity it would have to decay exponentially. It would therefore have to have negative energy, as usual for a bound state. There would therefore exist a wave function, namely the wave function of this bound eigenstate, for which the expectation value of the Hamiltonian was negative. The expectation value of the energy of a particle on a curved surface without any potential, however, is
\begin{equation}
\langle\hat{H}\rangle = \frac{\hbar^2}{2M}\sum_{i,j=1}^2 \int\!d^2r\,\sqrt{g}g^{ij}\partial_i\psi^*\partial_j\psi\;,
\end{equation}
where $g^{ij}$ is the contravariant metric tensor on the two-dimensional surface and $g$ is the determinant of its inverse matrix (the covariant metric). 
The integral is invariant under arbitrary coordinate changes, so we can show that the integrand is everywhere positive definite, for all $\psi(\mathbf{r}$, by transforming at any point to coordinates in which the metric tensor is $g^{ij}=\delta_{ij}$. No negative energy expectation values can exist, therefore---and therefore no negative eigenvalues and no bound states. Local curved regions can hold long-lived quasi-bound resonances, however, as we will see.

For potential-free cases in which an asymptotically flat plane contains a curved region, the quantum problem is thus essentially a scattering problem: how do incident plane waves from infinity propagate through the curved region? For the simple case of the sphere on a plane the quantum scattering problem can, like its classical counterpart, be solved exactly, albeit in the form of a Fourier series for the scattering amplitude in which the coefficients involve Legendre and Bessel functions.

\subsection{Time-independent Schr\"odinger equation for the sphere on a plane}
We consider a quantum particle of mass $M$ that moves in the geometry defined by the metric~\eqref{metric}, with sphere radius $R$ and contact angle $\alpha$ as before. 
In a quantum energy eigenstate of energy eigenvalue $E=\hbar^2 k^2/(2M)$, the wave function $\psi(\mathbf{r})$ of the particle obeys the time-independent Schr\"odinger equation
\begin{equation}\label{Schro1}
\frac{\hbar^2k^2}{2M} \Psi(\mathbf{r})= -\frac{\hbar^2}{2MR^2}\left[
\frac{\partial^2}{\partial \rho^2} + \frac{1}{2}  \frac{g'(\rho)}{g(\rho)}  \frac{\partial}{\partial \rho} +\frac{1}{g(\rho)} \frac{\partial^2}{\partial \theta^2} \right]\Psi,
\end{equation}
where $\mathbf{r}=(R\rho,\theta)$ in polar coordinates, 
with $g(\rho)$ again given by (\ref{grho}) and $g'(\rho) \equiv d g/d \rho$. We can then further decompose a generic energy eigenstate into simultaneous eigenstates of energy and angular momentum,
\begin{equation}
\Psi(\mathbf{r}) = \sum_m C_m e^{im\phi}\psi_m(\rho)\;,
\end{equation}
satisfying 
\begin{equation}\label{Schro2}
(kR)^2  \psi_m(\rho)= -\left[
\frac{\partial^2}{\partial \rho^2} + \frac{1}{2}  \frac{g'(\rho)}{g(\rho)}  \frac{\partial}{\partial \rho} -\frac{m^2}{g(\rho)} \right]\psi_m.
\end{equation}
We note that $\hbar m$ is the quantum analog of the classical angular momentum $J$, and that the wave function depends on $R$ and $k$ only through their product $kR$. 

For $\rho>\alpha$, (\ref{Schro2}) is just the usual Bessel equation for a radial eigenfunction in the plane, with the ordinary radial coordinate $r=R(\rho-\alpha+\sin\alpha)$ in the plane outside the sphere. So the normalized real solution for $\rho>\alpha$ can be written
\begin{align}\label{psq}
\psi_m(\rho) =& \cos(\delta_m)J_m\big(kR(\rho-\alpha+\sin\alpha)\big)\nonumber\\
&-\sin(\delta_m)Y_m\big(kR(\rho-\alpha+\sin\alpha)\big)\nonumber\\
\equiv & \mathrm{Re}\left(e^{i\delta_m}H_m\big(kR(\rho-\alpha+\sin\alpha)\big)\right)\;,
\end{align}
for some phase shift $\delta_m(kR)$, where $J_m$ and $Y_m$ are the Bessel and Neumann functions, respectively, and $H_m=J_m+iY_m$ is the Hankel function. For $\rho<\alpha$, however, (\ref{Schro2}) is the associated Legendre equation of order $\lambda$ such that $\lambda(\lambda+1) = (kR)^2$,
\begin{equation}
\lambda(kR) = \sqrt{(kR)^2+\frac{1}{4}}-\frac{1}{2}\;.
\end{equation}
This equation has only one solution that is regular at $\rho=0$, the associated Legendre function $P_{\lambda}^m(\cos\rho)$, and so we must have $\psi_m(\rho)=C_m P_\lambda^m(\cos\rho)$ for some real $C_m(kR)$.

The coefficients $C_m$ and $\delta_m$ are determined by imposing continuity of $\psi_m(\rho)$ and $\psi'_m(\rho)$ across the junction at $\rho=\alpha$ between the sphere and the plane. The result for the phase shift $\delta_m$ is
\begin{widetext}
\begin{equation}\label{PS}
\tan\delta_m = \frac{kR P^m_\lambda(\cos\alpha)J'_m(kR\sin\alpha)+\sin\alpha J_m(kR\sin\alpha) P^{'m}_\lambda(\cos\alpha)}{kRP^m_\lambda(\cos\alpha)Y'_m(kR\sin\alpha)+\sin\alpha Y_m(kR\sin\alpha)P^{'m}_\lambda(\cos\alpha)}\;,
\end{equation}
\end{widetext}
where the primes denote differentiation of each function with respect to its argument. In the two-dimensional version of standard partial wave scattering theory, the $\delta_m$ determine all scattering features, in the sense that an energy eigenstate of the scattering form
\begin{align}
\Psi_k(\mathbf{r}) &= e^{ikx}+\psi_{\mathrm{sc}}(\mathbf{r})\nonumber\\
\lim_{r\to\infty}\psi_{\mathrm{sc}}&=\sqrt{\frac{2}{\pi k r}}e^{i\frac{\pi}{4}}f(\theta)e^{ikr}
\end{align}
can be composed out of many $\psi_m$ as
\begin{equation}\label{psik}
\Psi_k(\mathbf{r}) = \sum_{m=-\infty}^\infty e^{i\delta_m}i^m e^{im\theta}\psi_m(\rho)\;.
\end{equation}
This follows from the asymptotic forms of the Bessel and Neumann functions and from the expansion of the two-dimensional plane wave in polar coordinates. The result by elementary algebra is just as for partial waves in three dimensions, 
\begin{equation}\label{dXs}
f(\theta) = \sum_{m=-\infty}^\infty \sin\delta_m e^{i\delta_m}e^{im\theta}\;,
\end{equation}
which implies the two-dimensional optical theorem
\begin{align}
\sigma :=&\lim_{r\to\infty} r \oint\!d\theta\, |\psi_{\mathrm{sc}}|^2 = \frac{2}{\pi k}\oint\!d\theta\,|f|^2\nonumber\\
=&\frac{4}{k}\sum_{m=-\infty}^\infty \sin^2\delta_m = \frac{4}{k}\mathrm{Im}[f(0)]\;.
\end{align}

We can therefore obtain exact differential cross-sections $d\sigma/d\theta = 2|f(\theta)|^2/(\pi k)$ for arbitrary $kR$ and $\alpha$, from the Fourier series (\ref{dXs}) for $f(\theta)$, with $\delta_m$ given in terms of Bessel, Neumann, and associated Legendre functions by (\ref{PS}). Using the asymptotic behaviors of the Bessel functions at small argument, and of the associated Legendre functions at small order $\lambda$, it is straightforward to show that for $kR\ll 1$, scattering becomes approximately isotropic ($f(\theta) = \sqrt{k\sigma/4}\times[1+\mathcal{O}(kR)^2]$), with
\begin{equation}
\lim_{kR \to0}\sigma=4\pi^2 k^3 R^4 \sin^8\frac{\alpha}{2}\;.
\end{equation}
For $\alpha\ll1$ as well as $kR\ll 1$, this agrees with the first Born approximation result of $\sigma = (\pi^2/64) k^3 R^4\sin^8\alpha$ which is obtained from Eqn.~(23) of Ref.~\cite{OMA18}, when we express our locally curved surface in the notation of Ref.~\cite{OMA18}. (The quantity $G(r)$ of Ref.~\cite{OMA18} becomes $G(r)\to (r/R)\theta(R\sin\alpha - r)$ in our case, $\theta(x)$ being the step function, and the parameters $\lambda_{1,2}$ of Ref.~\cite{OMA18} are both zero for us.) The first Born approximation of Ref.~\cite{OMA18} is only valid in our case when both $\alpha$ and $kR$ are small, so that the total effect of the curved region can be a small perturbation; the results of \cite{OMA18}, on the other hand, are valid for small local curvature deformations that are much more general in form than our sphere on a plane. 

\begin{figure}
\centering
\includegraphics[width=0.5\textwidth, trim=0mm 0mm 0mm 0mm ,clip]{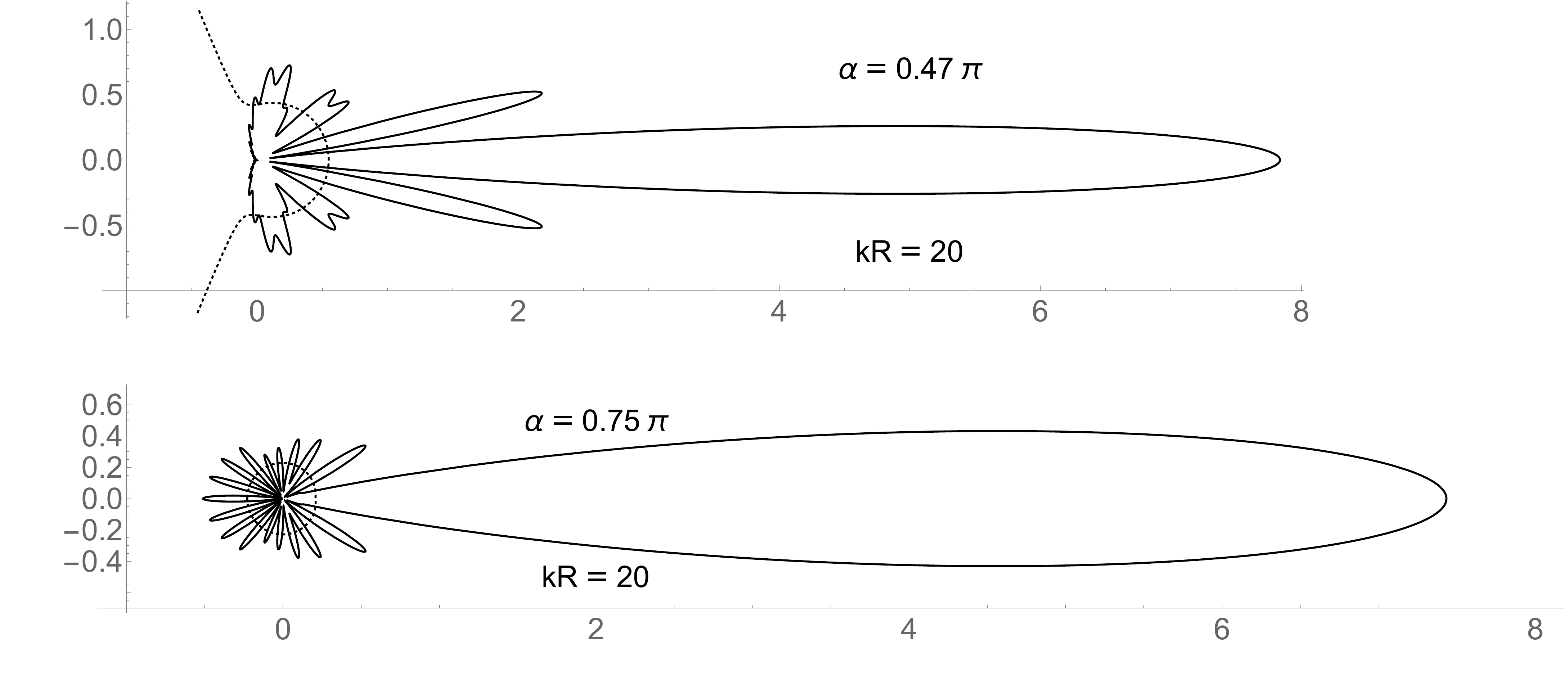}
\caption{Quantum differential cross sections divided by $R$, for $kR=20$. Although Cartesian axes are shown because the plots extend so far to the right, these are actually polar plots, with deflection angle $\theta=0$ corresponding to the horizontal-right direction, and radius from the origin at (0,0) giving $R^{-1}d\sigma/d\theta =2|f(\theta)|^2/(\pi kR)$, for scattering into the corresponding deflection angle. The corresponding classical differential cross sections divided by $R$, $R^{-1}db/d\theta$ as given by (\ref{classicalresult}), are superimposed as dashed contours. \textbf{Upper panel:} a slightly-less-than-hemisphere ($\alpha=0.47\pi$). \textbf{Lower panel:} a three-quarter sphere ($\alpha = 3\pi/4$). }
\label{spikefigs}
\end{figure}

Although at long wavelengths the isotropic quantum scattering is simpler than the caustics and cardioids of the classical differential cross section, at shorter wavelengths the angular dependence of quantum scattering becomes more complicated. See Fig.~\ref{spikefigs} for some examples, which illustrate most of the typical features for all $kR$ and $\alpha$; the classical differential cross sections in each case are shown in dots, for comparison. Once $kR$ is larger we can recognize that the quantum differential cross section mainly oscillates around the classical value, but beyond this there are significant differences.

At small deflection angles $\theta$ there is a dramatic spike of quantum scattering probability. In the limit of large $kR$, where one might expect from classical correspondence that the total quantum cross section would approach the classical value $2R\sin\alpha$, the narrow forward spike actually contributes a further $2R\sin\alpha$ in cross section all by itself, bringing the total quantum cross section to $4R\sin\alpha$ as $kR\to\infty$. As we will see in the next section, however, this large forward lobe in the differential cross-section is actually such a basic wave-mechanical scattering feature that it is not really due to the peculiar geometry of the sphere on a plane. Any round obstacle of the same size---for example, a hard disk---will produce a very similar spike of forward scattering for wavelengths much shorter than the obstacle size. The effect is a matter wave analogue of the optical Poisson spot, and while it is in principle a real phenomenon it is in practice difficult to observe because its existence depends on the incident particle beam having a coherence width broader than the target width (in this case $2R\sin\alpha$). At large $kR$, furthermore, where the forward spike is most dramatic, the spike becomes so narrowly concentrated around deflection angle zero that it could be difficult to distinguish it in experiments from the background of unscattered particles in the incident beam. There will usually be little reason for experimentalists to work hard to resolve the forward spike, moreover, since scattering experiments are usually intended to probe the internal structure of the scattering target, and the forward spike does not depend on this internal structure. In effect the forward spike is a dramatic wave-mechanical effect that is not actually very important.

The oscillations of $|f(\theta)|^2$ around the classical $db/d\theta$ persist at high $kR$; as we will see, they are a basic interference effect due to the presence of two classical trajectories for each scattering angle $\theta$. The classical caustics which occur for $ \alpha<\pi/2$ when these two trajectories merge are, as usual, softened by wave mechanics. The quantum scattering probability extends smoothly beyond these classical caustics; the appearance of quantum particles in a classically forbidden zone can be considered a form of tunnelling even though we do not have a potential barrier of the usual kind.

\begin{figure}
\centering
\includegraphics[width=0.45\textwidth, trim=0mm 0mm 0mm 0mm ,clip]{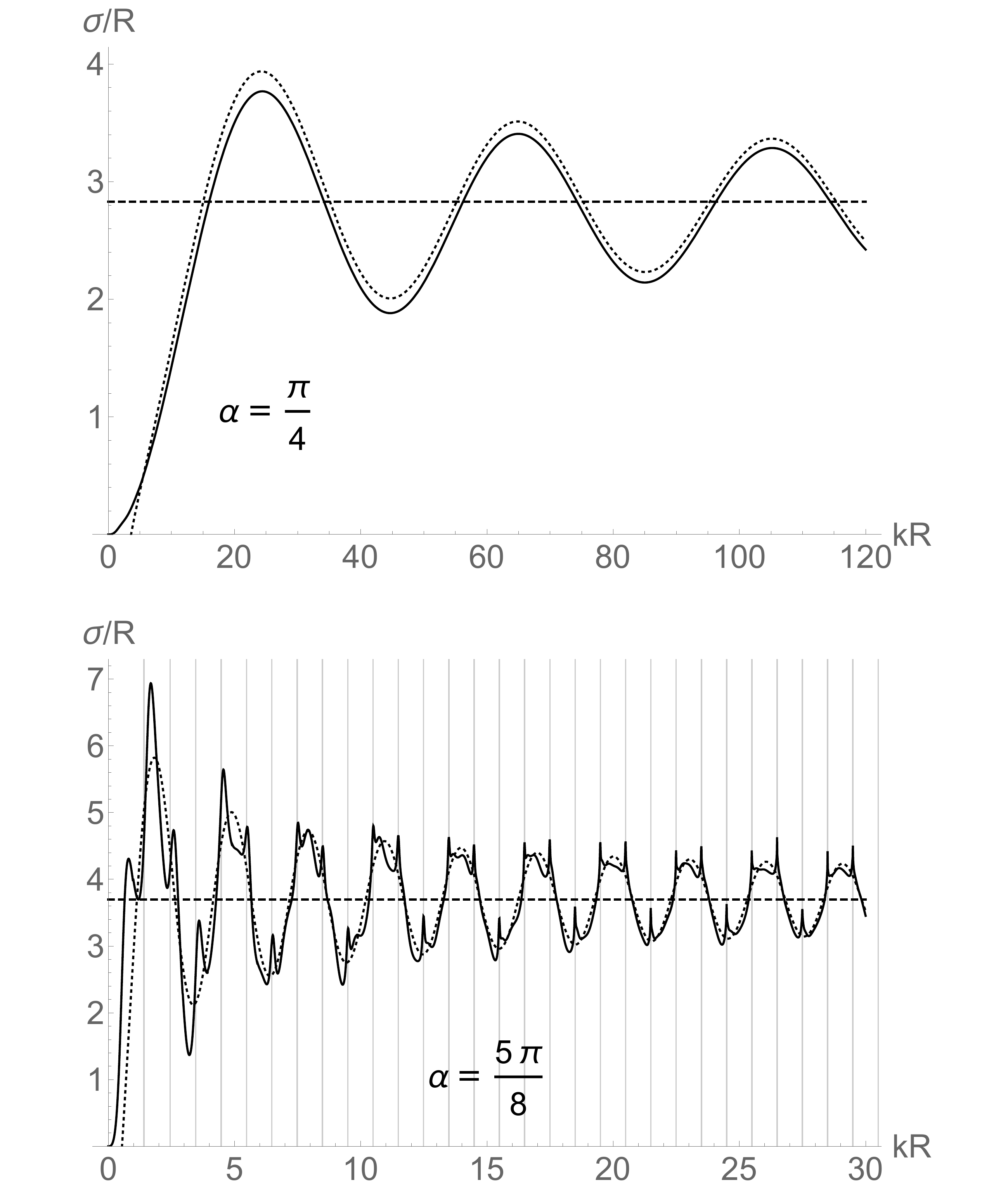}
\caption{Quantum total cross sections $\sigma$ divided by sphere radius $R$, as functions of $kR$, for a quarter-sphere ($\alpha=\pi/4$, upper panel) and a slightly-more-than-hemisphere ($\alpha=5\pi/8$, lower panel). The high-$kR$ limit $\sigma/R=4\sin\alpha$ is shown as a horizontal dashed line; the semi-classical approximation with tunnelling neglected, defined below in Section IV, is shown as a dotted curve. For sub-hemispheres like the case in the upper panel, the cross section oscillates as it approaches the asymptotic limit, following the semi-classical curve closely for $kR\gg1$ but with a small systematic deficit. As $\alpha$ approaches $\pi/2$ from below the oscillations at lower $kR$ become more complicated, but settle down to sinusoidal oscillation around $4\sin\alpha$ at higher $kR$. For $\alpha>\pi/2$ sharp peaks appear at all $kR = \sqrt{l(l+1)}$ for whole $l$ (vertical grid lines in the right plot). As $\alpha$ approaches $\pi$ and $kR$ increases, these peaks become narrower.}
\label{sigma1}
\end{figure}

\begin{figure}
\centering
\includegraphics[width=0.45\textwidth, trim=0mm 0mm 0mm 0mm ,clip]{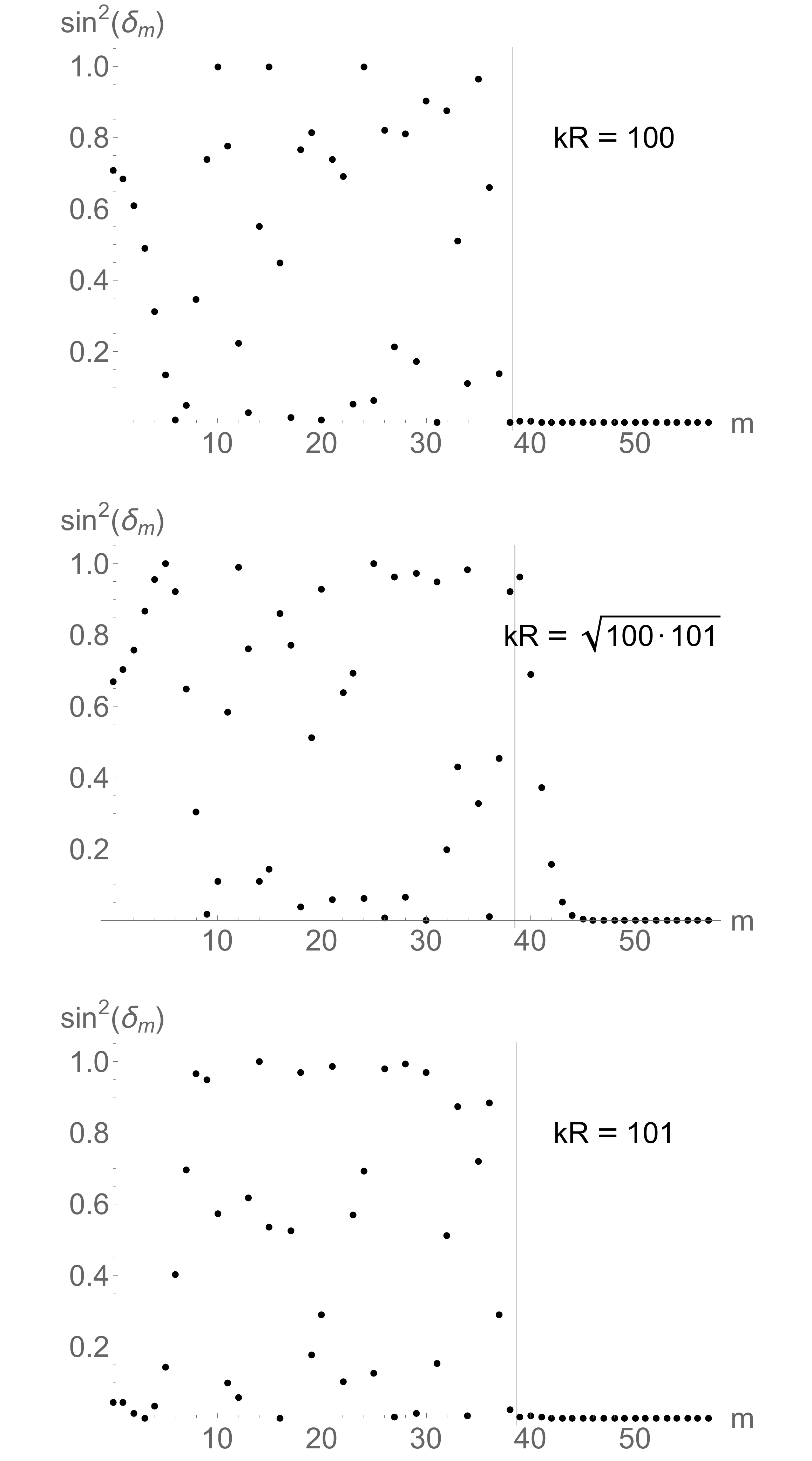}
\caption{Partial wave scattering probabilities $\sin^2\delta_m$ (vertical axes) as functions of angular quantum number $m$ (horizontal axes), for a nearly-full sphere ($\alpha=7\pi/8$), at $kR=100$ (top), $kR=101$ (bottom), and $kR=\sqrt{100\cdot 101}$ (middle). Vertical lines are at $m=kR\sin\alpha$. Similar plots for almost all $kR$ resemble the top and bottom panels qualitatively, with complicated jumbles of points that all fall to zero for $|m|>kR\sin\alpha$. Whenever $kR$ is close to $\sqrt{l(l+1)}$ for any whole number $l$, however, a few non-zero points extend past the usual $kR\sin\alpha$ limit. These anomalous points can in general be fitted quite well to a Gaussian. Similar patterns are seen for all $\alpha>\pi/2$.}
\label{triple}
\end{figure}

A further quantum effect which is not obvious in any individual plot of $|f(\theta)|^2$ shows up when we plot the total cross-section $\sigma$ as a function of $kR$, as in Fig.~\ref{sigma1}. For super-hemispherical cases $\alpha>\pi/2$ the cross section shows sharp resonance peaks whenever $kR=\sqrt{l(l+1)}$ for whole number $l$; these precise values correspond to the energy levels $\hbar^2 l(l+1)/(2MR^2)$ of a particle on an isolated full sphere of radius $R$, of which the corresponding eigenfunctions are the spherical harmonics $Y_{lm}(\rho,\theta)$. At higher $l$ and for $\alpha$ closer to $\pi$ the resonances become narrower and narrower, but their location does not change and they are present even for $\alpha$ only slightly greater than $\pi/2$. For sub-hemispherical cases $\alpha<\pi/2$, however, these peaks are completely absent.  A clue to what causes these sharp peaks can be seen when we examine the partial wave scattering probabilities $\sin^2(\delta_m)$, as shown in Fig.~\ref{triple}. For larger $kR$ these probabilities generally fall abruptly to (essentially) zero for $|m|>kR\sin\alpha$; this is because for $|m|$ beyond this limit the centrifugal barrier prevents particles with energy $\hbar^2k^2/(2M)$ from even reaching the contact circle at $r=R\sin\alpha$. When $k$ is close to $\sqrt{l(l+1)}$ for whole-number $l$, however, finite scattering probability extends somewhat past the $kR\sin\alpha$ limit, in an approximately half-Gaussian profile. The total cross section is simply the sum of all the partial wave probabilities, and so these additional non-negligible $\sin^2\delta_m$ contributions from $|m|>kR\sin\alpha$ are responsible for the sharp peaks in the cross section as a function of $kR$. As we will see in the next section, these contributions to scattering from angular momenta $kR\sin\alpha < |\hbar m| < kR$ are due to tunnelling into resonances related to the closed classical orbits within the sphere, when it overhangs the plane for $\alpha>\pi/2$. 

All of these features can be understood analytically by using the Wentzel-Kramers-Brillouin-Jeffreys semi-classical approximation to the quantum problem, at least qualitatively. Close quantitative agreement between the exact quantum treatment and the semi-classical approximation may require quite high $kR$, since the proportional errors in the semi-classical results are of order $(kR\sin\alpha)^{-1/2}$ in general, and (particularly for small $\alpha$) there can be unfortunately large $\alpha$-dependent pre-factors. At large enough $kR\sin\alpha$ to make these errors small, computing the exact differential cross section becomes numerically challenging; it also becomes a visual challenge to to compare  polar plots with very many radial fringes. In the next section we will therefore develop the semi-classical theory which becomes exact in the limit $kR\to\infty$, and show with numerical plots that for modest $kR\lesssim 100$ the agreement with the exact differential cross sections is already good enough to confirm the large-$kR$ trend.

\section{The Semi-Classical Problem}

For the conventional Schr\"odinger problem of motion in a potential, the semi-classical regime consists of cases in which the quantum wavelength is short compared to the scale on which the potential varies spatially. In this regime we apply the Wentzel-Kramers-Brillouin-Jeffreys (WKBJ) approximation to solve the time-independent Schr\"odinger equation for the quantum wave function using the method of characteristics. The wave function propagates simply along the characteristic curves in space, and these curves are a set of classical trajectories. Because of the need to find these classical trajectories as the first step towards solving the quantum problem, in general the WKBJ method is significantly more complicated in two or more dimensions than it is in the one-dimensional case that is more commonly presented in textbooks. Since our problem has exact rotational symmetry, however, we can reduce it to a one-dimensional radial problem before applying the WKBJ approximation. We will then see that for the sphere on a plane the WKBJ method provides a good approximation to exact quantum scattering whenever the wavelength is short compared to the contact radius $R\sin\alpha$ between the sphere and the plane.

\subsection{The WKBJ approximation}

The WKBJ approximation requires, first of all, that $kR$ be large. One then expands
\begin{equation}
\psi_m(\rho) = \exp\left[i\sum_{n=0}^\infty (kR)^{1-n}\int_{\rho_0}^\rho\!d\rho'\,K_{mn}(\rho')\right]
\end{equation}
and solves the dimensionless radial Schr\"odinger equation (\ref{Schro2}) order-by-order in $(kR)^{-1}$. The result is
\begin{align}\label{PsiSC}
\psi_m(\rho) =&  \left(g(\rho)-\frac{m^2}{(kR)^2}\right)^{-\frac{1}{4}}\sum_\pm A_\pm e^{\pm i kR\int_{\rho_0}^\rho\!d\rho'\,K_{m0}(\rho')}\nonumber\\
&+\mathcal{O}(kR)^{-1}
\end{align}
for
\begin{equation}
K_{m0}(\rho) = \sqrt{1-\frac{m^2}{(kR)^2g(\rho)}}\;,
\end{equation}
where as we recall from (\ref{grho}), $g(\rho)=\sin^2\rho$ for $\rho<\alpha$ (on the sphere), while for $\rho>\alpha$ we instead have $g(\rho)=(\rho-\alpha+\sin\alpha)^2$, which corresponds to $1/r^2$ for the usual radial coordinate $r=\rho-\alpha+\sin\alpha$ in the plane. 

In (\ref{PsiSC}) we have taken the lower limit of $\rho$ integration in the phase to be the turning point $\rho_0(kR/m)$ such that $g(\rho_0)=m^2/(kR)^2$. Choosing any other lower limit would simply shift the coefficients $A_\pm$, but choosing the turning point is convenient because it is at this point that the fourth-root prefactor in (\ref{PsiSC}) diverges and the WKBJ approximation breaks down. As usual we can interpolate through this breakdown region by approximating the radial Schr\"odinger equation there as an Airy equation. 

If we assume that for $\rho<\rho_0$, where (\ref{PsiSC}) is again valid, the wave function $\psi_m(\rho)$ consists only of the $\pm$ branch in (\ref{PsiSC}) that decays as $\rho$ decreases (evanescent wave in the classically forbidden region), then we obtain the condition $A_- = i A_+$. This assumption about the behavior of $\psi_m$ for $\rho<\rho_0$ will turn out to be valid except for a few special cases (the tunnelling resonances) that we will mention below. This means that for large $\rho\to\infty$ we will have
\begin{align}\label{PsiSCas}
\lim_{\rho\to\infty}\psi_m(\rho) &=2\frac{A_+e^{i\pi/4}}{\sqrt{r}}\mathrm{Re}\left(e^{-i\pi/4}e^{i kR\int_{\rho_0}^\rho\!d\rho'\,K_{m0}(\rho')}\right)\nonumber\\
\equiv & 2\frac{A_+e^{i\pi/4}}{\sqrt{r}} \mathrm{Re}\left(e^{i(\delta_m+\frac{\pi}{4})}e^{i k\int_{\frac{|m|}{k}}^r\!dr'\,\sqrt{1-\frac{m^2}{k^2r^2}}}\right)
\end{align}
for
\begin{equation}\label{deltamSC}
\delta_m = kR\int_{\rho_0}^\alpha\! d\rho\,\sqrt{1-\frac{m^2}{(kR)^2g(\rho)}} -k\int_{\frac{|m|}{k}}^r\!dr'\,\sqrt{1-\frac{m^2}{k^2r^2}}\;.
\end{equation}
Since 
\begin{equation}\label{Hankel}
\lim_{r\to\infty}H_m(kr) \propto \frac{1}{\sqrt{r}} e^{-i\pi/4}e^{i k\int_{\frac{|m|}{k}}^r\!dr'\,\sqrt{1-\frac{m^2}{k^2r^2}}}\;,
\end{equation}
we can recognize by comparison with (\ref{psq}) above that $\delta_m$ as given by (\ref{deltamSC}) is indeed the semiclassical result for the scattering phase shift. The integrals in (\ref{deltamSC}) can be performed explicitly, leading to our semiclassical expression for the scattering amplitude $f(\theta) = \sum_m e^{i\delta_m}e^{im\theta}\sin\delta_m$.

\begin{figure}
\centering
\includegraphics[width=0.45\textwidth, trim=0mm 0mm 0mm 0mm ,clip]{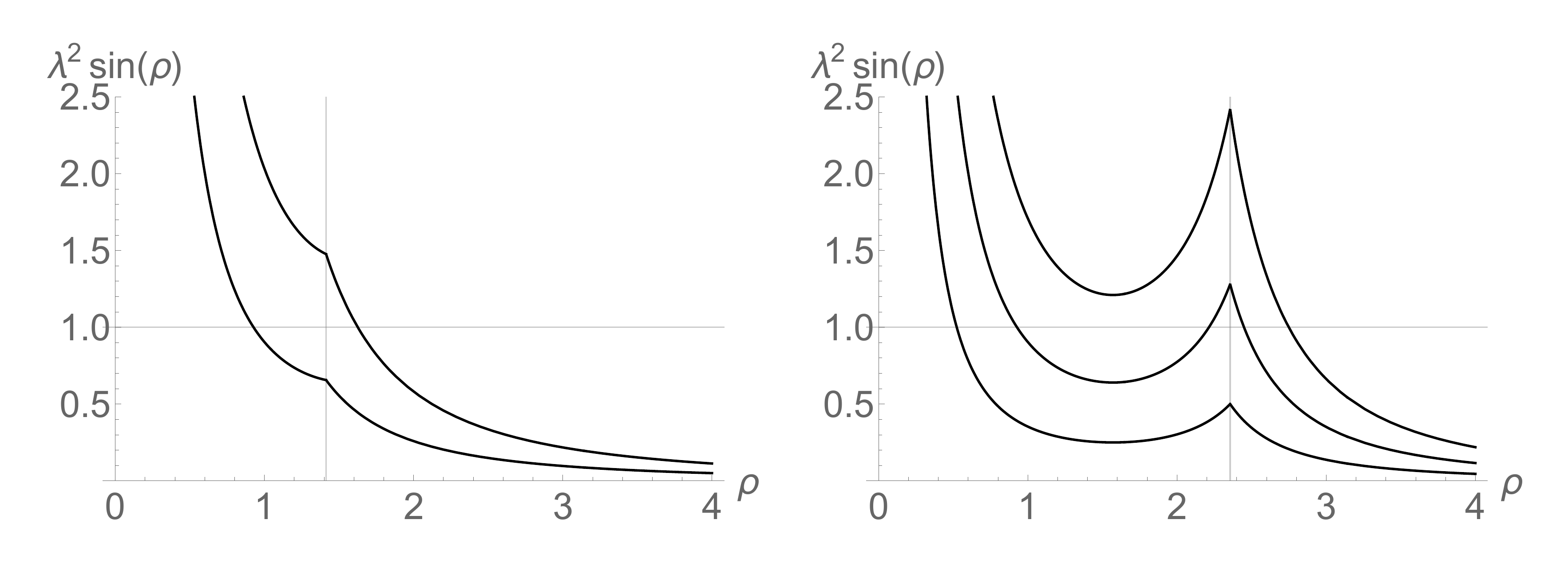}
\caption{The function $\lambda^2/g(\rho)$ for a sub-hemispherical and a super-hemispherical case (left panel, $\alpha=0.45\pi$; right panel, $\alpha=0.75\pi$), and for selected examples of $\lambda = m/(kR)$ (namely: 0.8 and 1.2 in the left panel; 0.5, 0.8, and 1.1 in the right panel). The intersection $\lambda^2/g(\rho)=1$ occurs at the turning point $\rho=\rho_0$. The vertical grid line marks $\rho=\alpha$ in both cases, where the form of $g(\rho)$ changes; for $\alpha>\pi/2$ the change in $g(\rho)$ gives the centrifugal barrier a maximum at $\rho=\alpha$, with an inner well at smaller $\rho$. In the sub-hemispherical left panel we see that for $|m|>kR\sin\alpha$ the turning point is outside the junction circle, $\rho_0>\alpha$, while for $|m|<kR\sin\alpha$ we instead have $\rho_0<\alpha$. The same is true for the outermost turning points in the super-hemispherical right panel, but within the range $kR\sin\alpha < |m|<kR$ there are classical orbits at energy $\hbar^2k^2/(2M)$ and angular momentum $\hbar m$ that remain inside the sphere ($\rho<\alpha$). Within this $m$ range the quantum particle can potentially tunnel through the centrifugal barrier, modifying the scattering behavior.}
\label{Turning}
\end{figure}

For $|m|>kR\sin\alpha$, (\ref{deltamSC}) implies immediately that $\delta_m=0$. We can see this by considering $\rho_0$ as a function of $m/(kR)$, as shown in Fig.~\ref{Turning}. What Fig.~\ref{Turning} shows is that for $|m|>kR\sin\alpha$ we have $\rho_0>\alpha$, so that the turning point in $\rho$ lies in the plane, fully outside the spherical extrusion; the centrifugal barrier turns the particle away before it reaches the contact circle at $r=R\sin\alpha$. For $|m|<kR\sin\alpha$, in contrast, the turning point $\rho_0<\alpha$ is on the spherical extrusion, where $g(\rho)$ represents the curved spherical surface. For $|m|>kR\sin\alpha$ and hence $\rho_0>\alpha$, therefore, we have $g(\rho)=(\rho-\alpha+\sin\alpha)^2 = r^{2}/R^2$ and the two integrals in (\ref{deltamSC}) cancel identically, leaving $\delta_m=0$ according to (\ref{deltamSC}).

This conclusion conflicts, however, with the $\delta_m\not=0$ for $|m|>kR\sin\alpha$ shown in the middle panel of Fig.~\ref{triple}, even though the $kR\doteq 100$ of Fig.~\ref{triple} should be large enough for semiclassical methods to work. To explain this discrepancy, we must now note the exceptions to our assumption of purely decaying $\psi_m(\rho)$ for $\rho<\rho_0$. As Fig.~\ref{Turning} shows, for $\alpha>\pi/2$ and $kR\sin\alpha < |m| < kR$ there are two additional turning points inside the outermost turning point $\rho_0$. For large $kR$ the tunnelling transmission amplitude through the centrifugal barrier peak at $\rho=\alpha$ is still exponentially small, and so for most $k$ these additional turning points will have negligible effect on scattering. For $(kR)^2 = l(l+1)$ for integer $l$, however, or very close, tunnelling through the centrifugal peak at $\rho=\alpha$ \emph{does} significantly change the phase relationship between $A_\pm$ in the $\rho>\rho_0$ region. This is clearly what is responsible for the anomalous $\delta_m \not=0$ for $|m|>kR\sin\alpha$ that were noted, for these special values of $k$ and only for $\alpha>\pi/2$, in Fig.~\ref{triple} above. We can therefore identify the sharp peaks in the cross section at these same values of $k$, which were noted in Fig.~\ref{sigma1} (again only for $\alpha>\pi/2$), as due to tunnelling through the centrifugal barrier peak into quasi-bound resonances with angular momentum $|m|>kR\sin\alpha$. These resonances correspond classically to closed great circle orbits on the spherical extrusion when it overhangs the plane for $\alpha>\pi/2$.

Although we are confident in this tunnelling explanation of the sharp scattering resonances at $k=\sqrt{l(l+1)}$ for $\alpha>\pi/2$, we will now ignore these sharp resonances and leave detailed study of these cases of tunnelling to future work. The project may be of some interest as an exercise in semiclassical technique, though for large $kR$ the resonances would likely be too narrow to observe experimentally, even if the rest of our model could be realized. We assume from now on that (\ref{deltamSC}) can be applied for all $kR$.

\subsection{The forward scattering spike}
This means that we take (\ref{deltamSC}) as correctly implying that $\delta_m = 0$ for $|m|>kR\sin\alpha$. This lets us simplify the semiclassical scattering amplitude $f_{SC}(\theta)$ by truncating the infinite sum over $m$ into a sum from $-\lfloor kR\sin\alpha\rfloor$ to $\lfloor kR\sin\alpha\rfloor$, where $\lfloor kR\sin\alpha\rfloor$ is the largest whole number not greater than $kR\sin\alpha$:
\begin{align}\label{fsc2}
f_{SC}(\theta) &=\sum_{m=-\lfloor kR\sin\alpha\rfloor}^{\lfloor kR\sin\alpha\rfloor} e^{im\theta} e^{i\delta_m}\sin\delta_m\nonumber\\
&= \frac{i}{2} \sum_{m=-\lfloor kR\sin\alpha\rfloor}^{\lfloor kR\sin\alpha\rfloor} e^{im\theta}\left(1-e^{2i\delta_m}\right)\nonumber\\
&\equiv \frac{i}{2}\frac{\sin\Big(\big(\lfloor kR\sin\alpha\rfloor+\frac{1}{2}\big)\theta\Big)}{\sin\frac{\theta}{2}} + \tilde{f}(\theta)\nonumber\\
\tilde{f}(\theta)&:=- \frac{i}{2}\sum_{m=-\lfloor kR\sin\alpha\rfloor}^{\lfloor kR\sin\alpha\rfloor}e^{i(2\delta_m + m\theta)}\;.
\end{align}
For large $kR\sin\alpha$, the first term in the third line of (\ref{fsc2}) is a sharp peak around $\theta=0$: it is responsible for the large forward spikes of scattering probability that we found in our exact quantum calculations, above. Because of the general form of partial wave scattering amplitudes $e^{i\delta_m}\sin\delta_m = (i/2)(1-e^{2i\delta_m})$, which is itself required by unitarity, this forward-spike term only depends on one feature of $\delta_m$---namely, that it vanishes for $|m|>kR\sin\alpha$. This means, as we mentioned above, that the same forward-scattering spike will appear for any scattering target for which $\delta_m$ has the same finite range of support in $m$. In the semi-classical limit this simply means that the scattering target has to have the same classical cross section $2R\sin\alpha$. 

\subsection{Classical paths}
The remaining $\tilde{f}(\theta)$ part of the scattering amplitude does depend on the particular form of $\delta_m$, as given semi-classically by (\ref{deltamSC}). The integrals in (\ref{deltamSC}) can be evaluated in closed form:
\begin{align}\label{deltamex}
\delta_m &= \frac{kR}{2}\Delta(\alpha,\frac{m}{kR})-\frac{m}{2}\Theta(\alpha,\frac{m}{kR})\nonumber\\
\Delta(\alpha,\mu)&:= 2\cos^{-1}\left(\frac{\cos\alpha}{\sqrt{1-\mu^2}}\right)-2\sqrt{\sin^2\alpha-\mu^2}\;,
\end{align}
and where $\Theta(\alpha,\mu)$ from (\ref{ThetaC}) is the classical scattering angle, as a function of sphere contact angle $\alpha$ and scaled angular momentum $\mu$. The path length difference $R\Delta(\alpha,\mu)$ is likewise the length of the portion of the classical path for $\alpha$ and $\mu$ which is shown thick in Fig.~\ref{classplot}, (\textit{i.e.} the length of the classical path between the tangents to the contact circle perpendicular to the incident and scattered directions), minus $2R\sin\alpha$ (which would be the distance between those two tangents, if there were no spherical extrusion and the particle just continued straight). We have stated (\ref{deltamex}) simply as the result of performing the integrals in (\ref{deltamSC}), which it is, but of course this is not a mere coincidence of integration. It can be derived from the classical mechanics for our system's Lagrangian, and in appropriately generalized form it holds for any rotationally symmetric scattering target.

Eqn.~(\ref{fsc2}) gives $\tilde{f}(\theta)$ as a sum of $\exp{\big(ikR\,S\big(m/(kR))\big)}$ over $m$, for 
\begin{align}\label{Sm}
S(\mu) &= 2\frac{\delta_{\mu kR}}{kR} + \mu \theta\nonumber\\
&= 2\cos^{-1}\left(\frac{\cos\alpha}{\sqrt{1-\mu^2}}\right)\nonumber\\
&\quad-2\sqrt{\sin^2\alpha-\mu^2}+\mu\big(\theta-\Theta(\alpha,\mu)\big)\;.
\end{align}
In the semi-classical limit of large $kR\sin\alpha$ it is accurate to approximate this sum as an integral \emph{and} also approximate the integral using the method of stationary phase. This approximation is accurate even though, for most of the terms in the sum, the change in the summand between one term and the next is \emph{not} small. In fact the approximation is accurate precisely because, for most terms in the sum, the summand rotates in phase between one term and the next rapidly and almost randomly: this means that most of the sum simply cancels. Around certain values of $m$, however, the phase of the summands changes only slowly with $m$, allowing the sum to be treated as a Riemann sum. Which values of $m$ it is, for which $kR\,S\big(m/(kR)\big)$ changes slowly, depends on $\theta$ and $\alpha$; the special values of $m$ are found by treating $\mu$ as if were a continuous variable, and finding the points at which $S(\mu)$ is stationary, $dS/d\mu = 0$.

As may again seem like a miraculous coincidence, differentiation of $S(\mu)$ with respect to $\mu$, including differentiating both $\Theta$ and $\Delta$, yields exactly the result that one would obtain if one forgot to differentiate $\Theta$ and $\Delta$:
\begin{align}
\frac{\partial}{\partial \mu}S(\mu)&= \theta - \Theta(\alpha,\mu)\quad\hbox{and so}\nonumber\\
\Theta\left(\alpha,\frac{m}{kR}\right) &\stackrel{!}{=}\theta\; \mathrm{mod}(2\pi)
\end{align}
determines the saddlepoints $m_\pm$.
This result is again no coincidence, however, but can be derived from the classical dynamics for any rotationally symmetric scattering target. Even the mod$(2\pi)$ here appears inevitably, because in $\tilde{f}(\theta)$ we are evaluating a discrete sum of phases, after all, and a phase which changes by nearly $2\pi$ between successive summand terms is equivalent to a slowly changing phase. The result is that the $\tilde{f}(\theta)$ sum (\ref{deltamSC}) is dominated by $m$ around precisely the classical angular momenta $J/\hbar$ which contribute to scattering into the angle $\theta$, as given by (\ref{thetabeta}). In particular we have two stationary points $\mu\to m_\pm kR$ for $m_\pm(\alpha,\theta)$:
\begin{equation}\label{mpm}
m_\pm = -kb_\pm(\alpha,\theta)
\end{equation}
as given by (\ref{btheta}) in Section II, above.

\subsection{Classical differential cross section}
Let us now examine the higher derivatives of $S(\mu)$ with respect to $\mu$, at $\mu \to m_\pm kR$. Since the classical impact parameters $b_\pm(\alpha,\theta)$ as given in (\ref{btheta}) are proportional to the sphere radius $R$, but independent of momentum, we can further see that the Taylor series of $S(m kR)$ around $m_\mp$ will be of the form
\begin{equation}
S\big((m_\pm + \Delta m)kR\big) = S(m_\pm kR) + \sum_{n=2}^\infty \frac{1}{n!} X_n(\alpha,\theta) \left(\frac{\Delta m}{kR}\right)^{n}
\end{equation}
for coefficients $X_n$ that do not depend on $kR$. This means that in the limit $kR\to\infty$, $\tilde{f}(\theta)$ in (\ref{deltamSC}) is indeed given by two Riemann sums for integrals with respect to $\mu = m/(kR)$, each of which describes the integral of a complex exponential, of which the phase has a large prefactor. We can therefore indeed apply the method of stationary phase to these integrals, approximating them as Gaussians, up to corrections that are smaller by factors of $1/(kR)$. 

In particular we note the second derivative of $S(\mu)$ at $\mu=b_\pm/R$,
\begin{align}
\frac{\partial^2}{\partial \mu^2} S(\mu) &\equiv \frac{\partial}{\partial \mu}\big(\theta-\Theta(\alpha,\mu)\big)\nonumber\\
&= R\frac{\partial\Theta}{\partial b_\pm}\equiv \left(\frac{1}{R}\frac{\partial b_\pm}{\partial\theta}\right)^{-1}\;.
\end{align}
Our stationary phase approximation to $\tilde{f}(\theta)$ is therefore
\begin{align}\label{paths2}
\tilde{f}(\theta) &\doteq -\frac{i}{2}\sum_\pm e^{ik\Delta(\alpha,\frac{m_\pm}{kR})} \sqrt{kR} \int_{-\infty}^\infty\!d\xi\,e^{i\frac{R}{2\partial_\theta b_\pm}\xi^2}\\
&= -i \sqrt{\frac{\pi k}{2}}\sum_\pm \sqrt{|\partial_\theta b_\pm|}\,e^{ikR\Delta\big(\alpha,-\frac{b_\pm(\alpha,\theta)}{R}\big)}e^{i\frac{\pi}{4}\mathrm{sgn}\,\partial_\theta b_\pm}\;.\nonumber 
\end{align}

Finally inserting (\ref{paths2}) into (\ref{fsc2}), and using $f(\theta)\to f_\mathrm{SC}(\theta)$ in the quantum differential cross section $d\sigma/d\theta = 2 |f(\theta)|^2/(\pi k)$, yields the semi-classical differential cross section as composed of three interfering terms:\begin{widetext}
\begin{equation}
\frac{d\sigma}{d\theta} = \left| \frac{\sin\Big(\big\lfloor kR\sin\alpha\rfloor +\frac{1}{2}\big)\theta\Big)}{\sqrt{2\pi k}\sin\frac{\theta}{2}}-\sum_\pm \sqrt{|\partial_\theta b_\pm|}\,e^{ikR\Delta(\alpha,-\frac{b_\pm}{R})}e^{i\frac{\pi}{4}\mathrm{sgn}\,\partial_\theta b_\pm}  \right|^2\;.
\end{equation}\end{widetext}
The incoherently adding ``diagonal'' terms in this semi-classical differential cross section consist of a sharp spike which approaches $2R\sin\alpha \delta(\theta)$ for $kR\to\infty$, plus exactly the classical differential cross section from Section II. The qualitatively non-classical interference terms, however, do not vanish for large $kR$; they simply oscillate more and more rapidly with $\theta$, providing many fine fringes. For large $kR$, the two terms with $\sqrt{|b_\pm|}$ factors dominate the differential cross section everywhere except at $\theta=0$, and these can be recognized as a kind of double-slit interference pattern, due to the two classical paths which exist for every scattering angle $\theta$ over the surface of the sphere on a plane.

\subsection{Comparison of exact and semi-classical differential cross sections}
\begin{figure}
\centering
\includegraphics[width=0.45\textwidth, trim=0mm 0mm 0mm 0mm ,clip]{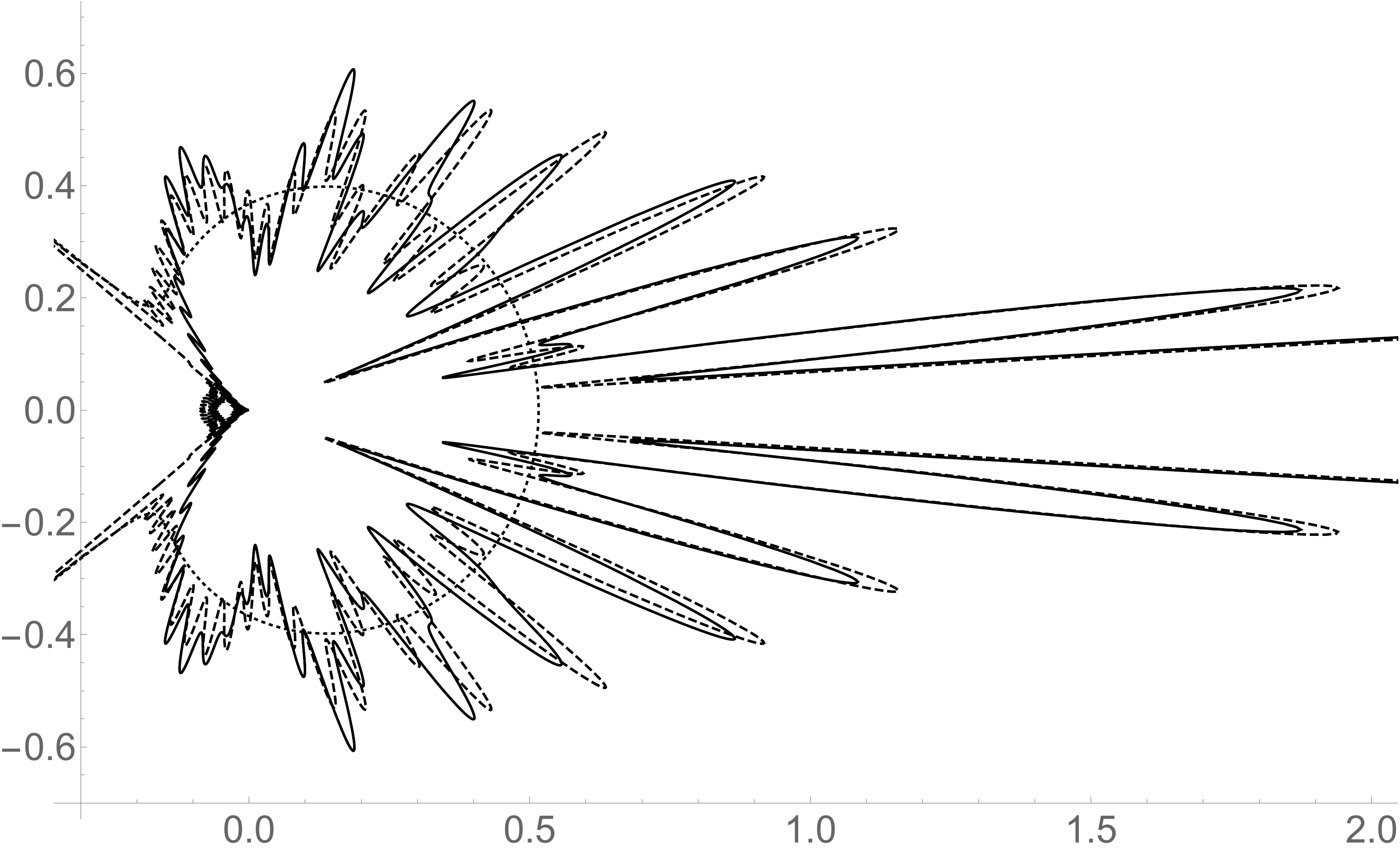}
\includegraphics[width=0.45\textwidth, trim=0mm 0mm 0mm 0mm ,clip]{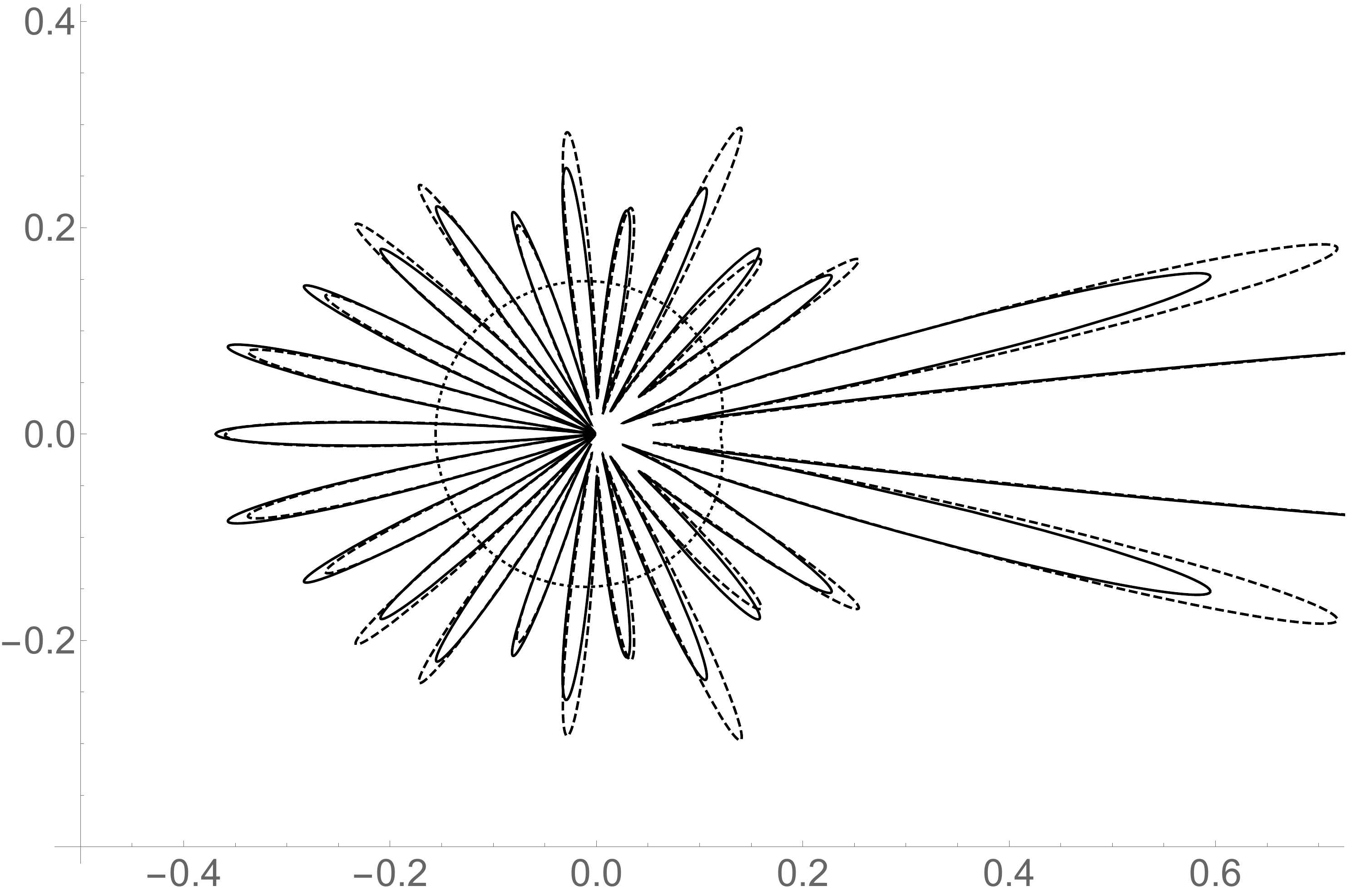}
\caption{Exact (solid) and semi-classical (dashed) differential cross sections (divided by $R$) for $kR=40$ and $\alpha=0.49\pi$ (upper plot), $\alpha=0.85\pi$ (lower plot). The classical differential cross sections are also shown (dotted) for comparison. The forward scattering spikes, suppressed to show detail for other angles, extend in the exact differential cross sections to 25.1 ($\alpha=0.49\pi)$ and to 3.66 ($\alpha=0.85\pi$), while the corresponding semi-classical values of $|f(0)|^2$ are 26.6 and 4.31. 
The semi-classical approximation already captures the angular spacing of the radial fringes well even at $kR\lesssim 20$, but higher $kR$ is required for the lengths of all the radial fringes to converge on the exact values, especially for $\alpha<\pi/2$. In particular the semi-classical approximation breaks down for $\alpha<\pi/2$ near and beyond the caustic $|\theta|=\theta_c$.}
\label{SC2}
\end{figure}

Figure \ref{SC2} compares the exact and sem-classical differential cross sections for two cases, both with $kR=40$, one with a (slightly) less than hemispherical extrusion ($\alpha=0.49\pi$), and one with an almost full sphere ($\alpha=0.85\pi$). It can be seen that the semi-classical approximation represents the angular positioning of the interference fringes quite well, even for this modestly large $kR$, though it does not always fully capture their amplitudes.

It can further be noted that the semi-classical approximation works better for the super-hemispherical case than for the sub-hemispherical one. There are two reasons for this. First of all the simple semi-classical method breaks down at the caustic at $\theta=\theta_c$, where the two $b_\pm$, and therefore the two $m_\pm$, coincide. Approximating the sum over $m$ in (\ref{fsc2}) as two separate Gaussian integrals is no longer a good approximation as we approach $\theta_c$. And for $|\theta|>\theta_c$ there are no stationary points $m_\pm$, because there are no classical trajectories, so the simple semi-classical method does not work at all. Improved semi-classical methods exist to resolve such problems\cite{Berry1}, but like the tunnelling resonances above, we leave them for future work.

Secondly, however, the semi-classical approximation works less well for $\alpha<\pi/2$ even for $|\theta|<\theta_c$, because although the extrema of $S(\mu)$ at $m_\pm kR$ are both nice maxima for all $\theta$, for $\alpha<\pi/2$ the extremum at $\mu = - b_+/R$ is, for most $|\theta|<\theta_c$, only a shallow dip within a steep slope. For sufficiently large $kR$ this still implies a deep minimum in the phase $kR\, S(\mu)$, but $kR$ has to be considerably larger to make the stationary phase approximation good around $m_+$ for $\alpha<\pi/2$ than it does for $\alpha>\pi/2$. Quantum corrections may therefore be needed for accurate differential cross sections at moderate $kR$ for sub-hemispherical extrusions.

\subsection{Exact and semi-classical total cross sections}
Using the optical theorem $\sigma(k) = (4/k)\mathrm{Im}f(0)$, we can conclude from (\ref{fsc2}) and (\ref{paths2}), with our results from Section 2 that $b'_+(\theta=0)=0$ and $b'_-(\theta=0)<0$, that in the semi-classical limit $kR\sin\alpha\gg 1 $ we have
\begin{align}
\sigma&\to R\Big(\tilde{\sigma}_\mathrm{SC}(\alpha,kR)+\mathcal{O}(kR)^{-1}\Big)\\
\tilde{\sigma}_\mathrm{SC}&=4\sin\alpha - \frac{2\sqrt{\pi\cot\frac{\alpha}{2}}}{\sqrt{kR}}\cos\big[2kR(\alpha-\sin\alpha)-\frac{\pi}{4}\big]\;.\nonumber
\end{align}
The dotted curves in Fig.~\ref{sigma1} are this $\tilde{\sigma}_\mathrm{SC}$. The agreement with the full quantum cross section is indeed good for $kR\sin\alpha\gg1$, except for the tunnelling resonance spikes for $\alpha>\pi/2$ that we have already discussed, and for a small but systematic discrepancy for $\alpha<\pi/2$, which we presume is due to the semi-classical failures at the classical caustics $\theta=\pm\theta_c$.

Apart from the problems that we have postponed to future work, concerning caustics for $\alpha<\pi/2$ and tunnelling resonances for $\alpha>\pi/2$, the semi-classical method clearly describes quantum scattering from the sphere on a plane quite well qualitatively for moderately large $kR$, with quantitative convergence as $kR\to0$. The striking non-classical features of quantum scattering are well explained semi-classically by the Poisson-spot-like forward spike and interference between the two classical trajectories that exist for each scattering angle $\theta$.

\section{Conclusions}
Our paper has been a technical exercise in theoretical physics, addressing a problem which is not only idealized but frankly exotic. We can hardly say that we have answered an urgent question; we must rather admit that our question is mainly interesting because of how thoroughly it can be answered. The same admission must be made for many solvable model problems, which nevertheless have their place in theoretical physics. Geodesic motion in curved spaces really is physically important, in the four-dimensional context of General Relativity; our two-dimensional example of piecewise uniform curvature may at least be of pedagogical value in providing some intuition about how curvature can affect motion.

More complicated two-dimensional cases than ours may also be of interest for understanding subtleties in the relationships between quantum and classical mechanics, possibly including subtleties of quantum chaos, because of the convenient billiard-like feature that the classical trajectories do not depend on energy, and so the entirety of classical dynamics can be visualized in two dimensions without loss of information, even though the phase space is actually four-dimensional. In our simple and symmetrical case, quantum-classical correspondence has turned out to be relatively straightforward, but interference and caustics and tunnelling have all appeared, even here. Further models of free particles on curved surfaces may be useful for theoretical studies of the quantum-classical frontier.

Experimental or even practical realizations of that general problem may not even be so far-fetched, moreover. Curved surfaces can be realized in graphene, for instance, and in such cases curvature will certainly influence band structure. The $V(x,y)$ potential effects which have assumed to be absent will surely be important in any such real two-dimensional problems; our results may nevertheless indicate an important feature of the interaction between curvature and potential energy, even though we have neglected the latter. We have observed that curvature alone cannot induce bound states, but as the right panel of our Figure \ref{Turning} shows, curvature can have dramatic effects on \emph{effective} potentials: it can significantly distort centrifugal barriers. If there is a potential well within a region of negative curvature, like our spherical extrusion, the softening of the centrifugal barrier due to curvature may allow more bound states to exist than one would expect without accounting for curvature. And we can conjecture, conversely, that if regions of positive curvature could somehow be realized, potential wells in these regions might possess fewer bound states.

JRA thanks Ali Mostafazadeh for drawing the nicely complementary paper \cite{OMA18} to his attention.

\end{document}